\documentclass[english,journal=jpcafh,manuscript=article,layout=twocolumn]{achemso}
\usepackage[T1]{fontenc}
\usepackage[latin9]{inputenc}
\usepackage{color}
\usepackage{babel}
\usepackage{refstyle}
\usepackage{amsmath}
\usepackage{amssymb}
\usepackage{graphicx}
\usepackage[numbers]{natbib}
\usepackage[unicode=true,pdfusetitle,
 bookmarks=true,bookmarksnumbered=false,bookmarksopen=false,
 breaklinks=false,pdfborder={0 0 1},backref=false,colorlinks=true]
 {hyperref}

\makeatletter

\title{Atomic Electronic Structure Calculations with Hermite Interpolating
Polynomials}

\author{Susi Lehtola}

\email{susi.lehtola@alumni.helsinki.fi}

\affiliation{Molecular Sciences Software Institute, Blacksburg, Virginia 24061,
United States}

\alsoaffiliation{Department of Chemistry, University of Helsinki, P.O. Box 55, FI-00014
University of Helsinki, Finland}

\AtBeginDocument{\providecommand\secref[1]{\ref{sec:#1}}}
\AtBeginDocument{\providecommand\subsecref[1]{\ref{subsec:#1}}}
\AtBeginDocument{\providecommand\eqref[1]{\ref{eq:#1}}}
\AtBeginDocument{\providecommand\figref[1]{\ref{fig:#1}}}
\AtBeginDocument{\providecommand\tabref[1]{\ref{tab:#1}}}
\providecommand{\tabularnewline}{\\}
\RS@ifundefined{subsecref}
  {\newref{subsec}{name = \RSsectxt}}
  {}
\RS@ifundefined{thmref}
  {\def\RSthmtxt{theorem~}\newref{thm}{name = \RSthmtxt}}
  {}
\RS@ifundefined{lemref}
  {\def\RSlemtxt{lemma~}\newref{lem}{name = \RSlemtxt}}
  {}

\usepackage[version=4]{mhchem}
\SectionNumbersOn

\setkeys{acs}{doi = true}

\@ifundefined{showcaptionsetup}{}{%
 \PassOptionsToPackage{caption=false}{subfig}}
\usepackage{subfig}
\makeatother

\begin{document}
\begin{abstract}
We have recently described the implementation of atomic electronic
structure calculations within the finite element method with numerical
radial basis functions of the form $\chi_{\mu}(r)=r^{-1}B_{\mu}(r)$,
where high-order Lagrange interpolating polynomials (LIPs) were used
as the shape functions $B_{\mu}(r)$. In this work, we discuss how
$\chi_{\mu}(r)$ can be evaluated in a stable manner at small $r$
and also revisit the choice of the shape functions $B_{\mu}(r)$.
Three kinds of shape functions are considered: in addition to the
$\mathcal{C}^{0}$ continuous LIPs, we consider the analytical implementation
of first-order Hermite interpolating polynomials (HIPs) that are $\mathcal{C}^{1}$
continuous, as well as numerical implementations of $n$-th order
($\mathcal{C}^{n}$ continuous) HIPs that are expressed in terms of
an underlying high-order LIP basis. Furnished with the new implementation,
we demonstrate that  the first-order HIPs are reliable even with large
numbers of nodes and that they also work with non-uniform element
grids, affording even better results in atomic electronic structure
calculations than LIPs with the same total number of basis functions.
We demonstrate that discontinuities can be observed in the spin-$\sigma$
local kinetic energy $\tau_{\sigma}$ in small LIP basis sets, while
HIP basis sets do not suffer from such issues; however, either set
can be used to reach the complete basis set limit with smooth $\tau_{\sigma}$.
Moreover, we discuss the implications of HIPs on calculations with
meta-GGA functionals with a number of recent meta-GGA functionals,
and find most Minnesota functionals to be ill-behaved. We also examine
the potential usefulness of the explicit control over the derivative
in HIPs for forming numerical atomic orbital basis sets, but find
that confining potentials are still likely a better option.
\end{abstract}
\global\long\def\ERI#1#2{(#1|#2)}%
\global\long\def\bra#1{\Bra{#1}}%
\global\long\def\ket#1{\Ket{#1}}%
\global\long\def\braket#1{\Braket{#1}}%
\global\long\def\erf#1{\text{erf }#1}%
\global\long\def\erfc#1{\text{erfc}\,#1}%
\global\long\def\sinc#1{\text{sinc\,}#1}%

\newcommand*\citeref[1]{ref. \citenum{#1}}
\newcommand*\citerefs[1]{refs. \citenum{#1}}
\newcommand*\HelFEM{\textsc{HelFEM}}
\newcommand\Erkale{\textsc{Erkale}}
\newcommand\Libxc{\textsc{Libxc}}
\newcommand\PsiFour{\textsc{Psi4}}
\newcommand\xtwodhf{\textsc{x2dhf}}
\newcommand\apcinfty{\mbox{aug-pc-$\infty$}}

\newref{subsec}{name = section~, names = sections~, Name = Section~, Names=Sections~}

%\bibnotesetup{ note-name = , use-sort-key = false }

\section{Introduction \label{sec:Introduction}}

In order to perform electronic structure calculations, the problem
needs to be discretized to fit a computer. The first step in electronic
structure theory is to determine the single particle states, usually
known as molecular orbitals, which are almost invariably expanded
in terms of analytic basis sets of a predefined form, as in the linear
combination of atomic orbitals (LCAO) approach, for example. However,
the goodness of the obtained solutions depends critically on the properties
of the basis set used for the expansion: if the basis set is a poor
fit to the problem, the results are not good either. 

Even though the bound solutions of the hydrogenic problem, $\hat{H}=-\nabla^{2}/2-Z/r$,
naïvely sound like a good starting point for finding a polyatomic
solution, such a basis is in fact a terrible starting point for electronic
structure problems, as the set of bound hydrogenic solutions must
be supplemented by the unbounded continuum states in order to form
a complete basis set. It has been known for a very long time that
the contribution from the continuum can be significant---comparable
in magnitude to that of the bound solutions---in many cases;\citep{Shull1955_JCP_1362,Ruffa1973_AJP_234}
this problem has been recently discussed for solutions of the hydrogenic
ground state problem for charge $Z'$ in the basis of the bound solutions
for $Z\neq Z'$ by \citet{Forestell2015_CJP_1009}. For related reasons,
the orbitals obtained from the one-electron part of the molecular
Hamiltonian $\hat{H}=-\nabla^{2}/2-\sum_{A}Z_{A}/r_{A}$ are likewise
a terrible guess for solving the self-consistent field (SCF) equations
occurring in both Hartree--Fock and Kohn--Sham\citep{Kohn1965_PR_1133}
theory.\citep{Lehtola2019_JCTC_1593}

Instead of the hydrogenic basis, most atomic-orbital calculations
employ basis sets of simpler analytic form, such as Gaussian type
orbitals and Slater type orbitals, the former of which have long dominated
quantum chemistry.\citep{Davidson1986_CR_681,Jensen2013_WIRCMS_273,Hill2013_IJQC_21}
The idea in both Gaussian and Slater type orbital basis sets is to
describe chemistry by analytic basis functions that ``look like''
atomic orbitals. The true atomic orbitals can be approached in such
a basis set given sufficiently many basis functions,\citep{Lehtola2019_IJQC_25968}
as the corresponding expansion coefficients are optimized to minimize
the total energy.\citep{Lehtola2020_M_1218} The benefit of this type
of approach is that a relatively compact atomic-orbital basis set
usually affords at least a qualitative level of accuracy for applications,
while larger basis sets can enable calculations that approach quantitative
accuracy with respect to experiments.\citep{Lehtola2019_IJQC_25968} 

However, the accuracy of analytic basis sets is limited, and in the
case of Gaussian basis sets, basis set truncation errors in the order
of 1 m$E_{h}$ are typically observed in total energies of heavy atoms.\citep{Malli1993_PRA_143,Koga2000_CPL_473,Lehtola2020_JCP_134108}
An alternative to employing analytical basis sets that approximate
the true form of atomic radial functions is to switch to methods that
use exact radial functions. The exact radial functions can be solved
with fully numerical methods; such calculations have been recently
reviewed in \citeref{Lehtola2019_IJQC_25968}. 

The general idea in fully numerical methods is to forego the chemical
intuition inherent in the LCAO approach, and instead directly solve
the differential equations arising from the Schrödinger equation for
the unknown orbitals. In the case of single atoms, the problem reduces
to the determination of the atomic radial functions, and yields numerical
atomic orbitals (NAOs). All fully numerical methods can be systematically
made more accurate, which allows the determination of SCF total energies
directly at the complete basis set (CBS) limit.

The finite difference method (FDM) is the traditional method of choice
in electronic structure, and it has been employed in a number of density
functional implementations for atoms.\citep{Lehtola2019_IJQC_25968}
However, the method of choice for solving differential equations of
various forms across various disciplines is not the FDM but the finite
element method (FEM). The great benefit of FEM is that it is a variational
method unlike FDM, and that it is straightforward to tailor the numerical
basis functions used in FEM to optimize the cost and accuracy of the
solution. FEM has been applied to many kinds of problems in the quantum
chemistry literature.\citep{Lehtola2019_IJQC_25968} 

We have recently published a free and open source\citep{Lehtola2022_WIRCMS_1610}
program for finite element calculations on atoms\citep{Lehtola2019_IJQC_25945,Lehtola2020_PRA_12516}
and diatomic molecules\citep{Lehtola2019_IJQC_25944,Lehtola2020_MP_1597989}
called \HelFEM{}.\citep{Lehtola2018__} \HelFEM{} affords an easy
way to approach CBS limit total energies for density functionals,
for example, as the total energies are computed variationally within
each numerical basis set, and the numerical basis sets can be systematically
extended towards the CBS limit. In contrast, the traditionally used
FDM does not satisfy the variational theorem and can give estimated
total energies that are above or below the CBS value.

We have shown that the FEM approach used in \HelFEM{} routinely affords
sub-$\mu E_{h}$ total energies for Hartree--Fock (HF) and density
functional calculations with local density approximations (LDAs),
generalized gradient approximations (GGAs) as well as meta-GGA functionals.\citep{Lehtola2019_IJQC_25945,Lehtola2019_IJQC_25944,Lehtola2020_PRA_32504,Lehtola2020_PRA_12516,Lehtola2020_MP_1597989,Lehtola2020_JCP_144105,Lehtola2020_JCP_134108,Lehtola2021_JCTC_943}
We have also shown that range-separated functionals can be implemented
within the same scheme.\citep{Lehtola2020_PRA_12516} 

Although some atomic FDM solvers that can also handle meta-GGA functionals,
such as the Atomic Pseudopotentials Engine (APE),\citep{Oliveira2008_CPC_524}
have been reported in the literature, \HelFEM{} was to the best of
our knowledge the first FEM program to be able to perform such calculations.

So far, all of our work has employed Lagrange interpolating polynomial
(LIP) shape functions for FEM, yielding a $\mathcal{C}^{0}$ continuous
numerical basis set. In this work, we will reinvestigate the choice
of the shape functions. In addition to LIPs, we will consider an analytical
implementation of first-order Hermite interpolating polynomials (HIPs),
and show that it affords high accuracy with compact expansions through
the use of high-order polynomial basis functions and a non-uniform
element grid. We find that the HIP basis affords lower energies than
a LIP basis with the same total number of basis functions, and thereby
can recommend the use of first-order HIPs instead of LIPs. 

Moreover, we will also consider a numerical implementation of higher-order
HIP basis functions solved in terms of a LIP basis with a large number
of nodes. Furnished with this implementation, we demonstrate that
discontinuities in the two meta-GGA ingredients---local kinetic energy
density $\tau$ and density Laplacian $\nabla^{2}n$---can be observed
in HF calculations employing small numerical basis sets not yet converged
to the CBS limit. In specific, the $\mathcal{C}^{0}$ LIP basis yields
discontinuous $\tau$ and $\nabla^{2}n$, the $C^{\text{1}}$ HIP
basis yields a cuspy $\tau$ and discontinuous $\nabla^{2}n$, while
the $\mathcal{C}^{2}$ second-order HIP basis yields a smooth $\tau$
and cuspy $\nabla^{2}n$.

However, we also demonstrate that when a large enough numerical basis
set that reproduces the CBS limit is used, both $\tau$ and $\nabla^{2}n$
come out smooth from both HF calculations as well as calculations
with $\tau$-dependent meta-GGA functionals.

The layout of this work is as follows. Next, in \secref{Theory},
we will discuss the theory of FEM (\subsecref{Finite-element-method})
and the choice of the basis polynomials used as the shape functions
(\subsecref{Basis-Polynomials}). The results are discussed in \secref{Results}.
We discuss how numerical instabilities in the evaluation of the numerical
basis functions near the nucleus can be avoided through the use of
high-order Taylor expansions (\subsecref{numstab}), show that HIPs
afford excellent results with non-uniform grids and large numbers
of nodes and overperform LIPs (\subsecref{suphip}), analyze the effect
of HIPs in calculations with a selection of meta-GGA density functional
approximations (\subsecref{metaGGA}), and examine their potential
benefits for building NAO basis sets (\subsecref{nao}). The article
concludes in a summary and discussion in \secref{Summary-and-Discussion}.
Atomic units are used throughout, unless specified otherwise.

\section{Theory \label{sec:Theory}}

\subsection{Finite Element Method \label{subsec:Finite-element-method}}

The employed finite element formalism has been extensively discussed
in \citerefs{Lehtola2019_IJQC_25945} and \citenum{RamMohan2002__},
but will be briefly summarized for completeness as we revisit the
choice of the shape functions in this work. The radial functions for
spin $\sigma$ are expanded in the numerical basis set as

\begin{equation}
R_{\sigma nl}(r)=\sum_{\mu}C_{\mu n}^{\sigma(l)}\chi_{\mu}(r),\label{eq:orb}
\end{equation}
where $\chi_{\mu}(r)$ are the numerical basis functions, which in
turn are given by 
\begin{equation}
\chi_{\mu}(r)=B_{\mu}(r)/r.\label{eq:fem-func}
\end{equation}
The shape functions $B_{\mu}(r)$ in \eqref{fem-func} are piecewise
polynomials defined in terms of $N_{\text{elem}}$ \emph{elements},
each element $i$ ranging from $r=r_{i}$ to $r=r_{i+1}$; the shape
functions therefore have finite support. Note that in order to avoid
\eqref{fem-func} diverging at the nucleus, all shape functions are
required to vanish at the origin, $B_{\mu}(r)\to0$ for $r\to0$,
and this is accomplished by removing the shape function describing
the function value at the nucleus.\citep{Lehtola2019_IJQC_25945}
Because of this removal, all the remaining numerical basis functions
become well defined with $\chi_{\mu}(r)\to B_{\mu}'(0)$ when $r\to0$.

The finite element grid defines the values $\{r_{i}\}$; the ``exponential
grid'' of \citeref{Lehtola2019_IJQC_25945} 
\begin{equation}
r_{i}=(1+r_{\infty})^{i^{z}/N_{\text{elem}}^{z}}-1\label{eq:expgrid}
\end{equation}
is used in the present work. The parameter $r_{\infty}$ in \eqref{expgrid}
is the \emph{practical infinity} employed in the calculation, beyond
which all orbitals vanish; the default value\citep{Lehtola2019_IJQC_25945}
$r_{\infty}=40a_{0}$ is used for all calculations in this work. The
parameter $z$ in \eqref{expgrid} controls the structure of the element
grid: the larger $z$ is, the more points are placed close to the
nucleus. The value $z=2$ has been found to afford results with excellent
accuracy in HF calculations,\citep{Lehtola2019_IJQC_25945} and the
value $z=2$ is also used in the present work unless specified otherwise.

\subsection{Basis Polynomials \label{subsec:Basis-Polynomials}}

Three types of shape functions will be investigated: Lagrange interpolating
polynomials (LIPs), as well as Hermite interpolating polynomials (HIPs)
of the first and second orders. The shape functions are expressed
within each element $r\in[r_{i},r_{i+1}]$ in terms of a primitive
coordinate $x\in[-1,1]$ obtained with the transformation
\begin{equation}
x(r)=2\frac{r-r_{i}}{r_{i+1}-r_{i}}-1,\ r\in[r_{i},r_{i+1}].\label{eq:coordtrans}
\end{equation}
Integrals are computed over the elements by quadrature with $N_{\text{quad}}$
points, and all calculations are converged to the quadrature limit.
A Chebyshev quadrature rule transformed to unit weight factor is employed
for this purpose, as it provides nodes and weights in easily computable
analytical form;\citep{PerezJorda1994_JCP_6520} note that the rule
does not place any points at the element boundaries.

\subsubsection{Lagrange Interpolating Polynomials \label{subsec:LIP}}

The LIP basis was chosen as the default in \HelFEM{} in previous
works.\citep{Lehtola2019_IJQC_25944,Lehtola2019_IJQC_25945} The LIP
basis is defined by a set of nodes $\{x_{i}\}_{i=1}^{N_{\text{nodes}}}$
satisfying $x_{1}=-1<x_{2}<\dots<x_{N_{\text{nodes}}}=1$ as

\begin{equation}
L_{i}(x)=\prod_{\begin{array}{c}
j=1\\
j\neq i
\end{array}}^{N_{\text{nodes}}}\frac{x-x_{j}}{x_{i}-x_{j}}.\label{eq:lip}
\end{equation}
LIPs satisfy the important property 
\begin{equation}
L_{i}(x_{j})=\delta_{ij}\label{eq:lipsys}
\end{equation}
Because of \eqref{lipsys}, the coefficient of the $i$th LIP $L_{i}(x)$
in the expansion of any given function $f(x)$
\begin{equation}
f(x)=\sum_{i}c_{i}L_{i}(x)\label{eq:lip-exp}
\end{equation}
is simply given by the value of the function at the corresponding
node $f(x_{i})$
\begin{equation}
f(x)=\sum_{i=1}^{N_{\text{nodes}}}f(x_{i})L_{i}(x),\label{eq:rad-lip}
\end{equation}
as is easily seen by evaluating \eqref{lip-exp} at $x=x_{j}$.

Let us now consider the continuity of the representation across element
boundaries. The continuity is guaranteed, since the right-most node
$x_{N_{\text{nodes}}}=1$ of the left-hand element coincides with
the left-most node $x_{1}=-1$ of the right-hand element; these two
LIPs are thus identified as pieces of the same shape function.

Boundary conditions at the nucleus and at infinity are handled by
removing the first and last numerical basis function, respectively;
this ensures that the wave function vanishes at $r_{\infty}$ and
that $B_{\mu}(r)/r$ does not diverge at the nucleus.\citep{Lehtola2019_IJQC_25945}

The Runge instability that arises for large numbers of uniformly placed
nodes is avoided by choosing the nodes with the Gauss--Lobatto quadrature
formula. This enables the use of extremely high-order LIPs with favorable
accuracy properties,\citep{Lehtola2019_IJQC_25945} as will also be
demonstrated later in this work; 15-node LIPs were chosen as the default
in \citeref{Lehtola2019_IJQC_25945}.

Although LIPs are only $\mathcal{C}^{0}$ functions and thereby do
not guarantee derivatives to be continuous across element boundaries,
we have demonstrated in a variety of studies performed at the density
functional and HF levels of theory\citep{Lehtola2021_JCTC_943,Lehtola2020_PRA_32504,Lehtola2020_PRA_12516,Lehtola2020_MP_1597989,Lehtola2020_JCP_144105,Lehtola2020_JCP_134108,Lehtola2019_IJQC_25945,Lehtola2019_IJQC_25944}
that the total energy converges smoothly to the complete basis set
limit when more elements are added in the calculation. The rationale
for this behavior is that the kinetic energy term in the Hamiltonian
imposes penalties on discontinuous derivatives across boundaries.\citep{Lehtola2019_IJQC_25945}
We will investigate discontinuities across element boundaries in detail
in \subsecref{metaGGA}.

\subsubsection{Analytic First-order Hermite Interpolating Polynomials \label{subsec:HIP}}

First-order HIPs, which are $\mathcal{C}^{1}$ continuous and explicitly
guarantee the continuity of the first derivative across element boundaries,
can be expressed in terms of LIPs as
\begin{align}
H_{2i-1}(x) & =h_{i}(x),\ H_{2i}=\mathfrak{h}_{i}(x)\label{eq:hip}\\
h_{i}(x) & =\left[1-2(x-x_{i})L_{i}'(x_{i})\right]\left[L_{i}(x)\right]^{2}\label{eq:hip-func}\\
\mathfrak{h}_{i}(x) & =(x-x_{i})\left[L_{i}(x)\right]^{2}\label{eq:hip-deriv}
\end{align}
It is easy to see that these functions satisfy the properties 
\begin{equation}
\begin{array}{cc}
h_{i}(x_{j})=\delta_{ij} & \mathfrak{h}_{i}(x_{j})=0\\
h_{i}'(x_{j})=0 & \mathfrak{h}_{i}'(x_{j})=\delta_{ij}
\end{array}\label{eq:hsys}
\end{equation}
As can be seen from \eqref{hsys} in analogy to the discussion on
LIPs and \eqref{lipsys} in \subsecref{LIP}, odd-numbered functions
($h_{i}$) carry information on the function value at the corresponding
node, only, while the even-numbered functions ($\mathfrak{h}_{i}$)
carry information only on the derivative at the corresponding node
\begin{equation}
f(x)=\sum_{i}\left[f(x_{i})h_{i}(x)+f'(x_{i})\mathfrak{h}_{i}(x)\right].\label{eq:rad-hip}
\end{equation}

Nodes at the element boundaries are again used to glue the shape functions
representing function values or derivatives together at each side
of the boundary. The first $h_{i}$ is removed to satisfy the boundary
condition at the nucleus, while the first $\mathfrak{h}_{i}$ describes
the electron density at the nucleus.\citep{Lehtola2019_IJQC_25945}
To satisfy the boundary condition at the practical infinity $r_{\infty}$,
the last $h_{i}$ is removed such that the radial function vanishes
at $r=r_{\infty}$. If the last $\mathfrak{h}_{i}$ is removed, as
well, then the radial function's derivative will also vanish at $r=r_{\infty}$;
see \subsecref{nao} for a case study.

Although HIPs were already examined in \citeref{Lehtola2019_IJQC_25945},
the work had two deficiencies. First, in contrast to this work (\eqref{hip}),
the implementation of \citeref{Lehtola2019_IJQC_25945} employed a
primitive polynomial representation ($x^{i}$ instead of $L_{i}(x)$)
and thereby is numerically unstable for large numbers of nodes. Second,
and more importantly, the matching across element boundaries did not
correctly take into account the ramifications of non-uniform elements:
the coordinate scaling $r\to Lr$ in \eqref{coordtrans} that results
in derivatives scaling as ${\rm d}^{n}/{\rm d}r^{n}\to L^{-n}{\rm d}^{n}/{\rm d}r^{n}$
was not taken into account in the implementation, causing the poor
results observed for non-uniform grids. Note that in order to make
the derivatives match on the boundaries, one must scale the basis
functions describing $n$:th derivatives at the nodes $f_{i}^{(n)}(x_{j})=\delta_{ij}$
by $L^{n}$ to obtain $f^{(n)}(r)=1$ at the element boundaries. As
will be shown in \subsecref{suphip}, HIPs in fact afford better accuracy
than LIPs with the same total number of basis functions also when
non-uniform grids are employed, although either basis can be used
to reach the CBS limit (\subsecref{metaGGA, nao}).

\subsubsection{Numerical Hermite Interpolating Polynomials \label{subsec:numHIP}}

For comparison and simplicity, we have also implemented higher-order
HIPs numerically. The HIPs can be solved numerically in terms of LIPs
from the linear system of equations corresponding to generalizations
of \eqref{rad-hip} to higher orders. For instance, the expansion
for the second order reads
\begin{equation}
f(x)=\sum_{i}[f(x_{i})h_{i}(x)+f'(x_{i})\mathfrak{h}_{i}(x)+f''(x_{i})\tilde{\mathfrak{h}}_{i}(x)]\label{eq:hip2}
\end{equation}
and the functions $h_{i}(x)$, $\mathfrak{h}_{i}(x)$ and $\tilde{\mathfrak{h}}_{i}(x)$
satisfy the equations

\begin{equation}
\begin{array}{ccc}
h_{i}(x_{j})=\delta_{ij} & \mathfrak{h}_{i}(x_{j})=0 & \tilde{\mathfrak{h}}_{i}(x_{j})=0\\
h_{i}'(x_{j})=0 & \mathfrak{h}_{i}'(x_{j})=\delta_{ij} & \,\tilde{\mathfrak{h}}_{i}'(x_{j})=0\\
h_{i}''(x_{j})=0 & \mathfrak{h}_{i}''(x_{j})=0 & \,\tilde{\mathfrak{h}}_{i}''(x_{j})=\delta_{ij}
\end{array}\label{eq:h2sys}
\end{equation}
where $i$ and $j$ are node indices, $i,j\in[1,N_{\text{nodes}}^{\text{HIP}}]$.
In our implementation, the general $n$:th order HIP basis functions
(for example, \eqref{hip2} for $n=2$), which guarantees $\mathcal{C}^{n}$
continuity, is re-expressed in terms of LIPs with $N_{\text{nodes}}^{\text{LIP}}=(n+1)N_{\text{nodes}}^{\text{HIP}}$
as $\boldsymbol{H}(x)=\boldsymbol{L}(x)\boldsymbol{T}$, where $\boldsymbol{H}$
are the HIP basis functions, $\boldsymbol{L}$ are the underlying
LIP functions, and $\boldsymbol{T}$ is the transformation matrix.
Gauss--Lobatto nodes are used for both the $N_{\text{nodes}}^{\text{HIP}}$
HIP nodes as well as the $N_{\text{nodes}}^{\text{LIP}}$ LIP nodes,
as discussed in \subsecref{LIP}. The transformation matrix $\boldsymbol{T}$
is solved numerically by inverting \eqref{h2sys}. 

\section{Results \label{sec:Results}}

\subsection{Numerically Stable Evaluation of Basis Functions \label{subsec:numstab}}

Before proceeding with electronic structure calculations, it is worthwhile
to discuss the stable evaluation of the numerical basis functions
in regions close to the nucleus. Exploratory calculations performed
as part of this work showed that in some cases, calculations with
otherwise well-behaved meta-GGA functionals failed to reach SCF convergence
and showed extremely large orbital gradients. We found the difference
between well-behaved and ill-behaved calculations to center around
the few closest quadrature points to the nucleus that had extremely
large values for $\tau$, and were able to remove this issue by reformulating
the numerical basis functions in a more stable manner.

Taylor expanding the basis functions of \eqref{fem-func} around $r=0$
shows that they are in principle well-behaved everywhere as
\begin{align}
\chi_{\mu}(r) & =\frac{B_{\mu}(r)}{r}=\frac{B_{\mu}(0)+B_{\mu}'(0)r+\frac{1}{2}B_{\mu}''(0)r^{2}+\dots}{r}\nonumber \\
 & =B_{\mu}'(0)+\frac{1}{2}B_{\mu}''(0)r+\dots\label{eq:chi-taylor}
\end{align}
where $B_{\mu}(0)=0$,\citep{Lehtola2019_IJQC_25945} even though
the evaluation of $B_{\mu}(r)/r$ and its derivatives is unstable
for small $r$. For this reason, we employ Taylor expansions to evaluate
$\chi_{\mu}(r)$ and its derivatives close to the nucleus, that is,
for $r<R$. 

Since all matrix elements are evaluated by quadrature, the use of
\eqref{chi-taylor}\emph{ is not an approximation}. Instead, \eqref{chi-taylor}
amounts to a redefinition of the numerical basis functions near the
nucleus
\begin{equation}
\chi_{\mu}(r)=\begin{cases}
B_{\mu}(r)/r, & r\geq R\\
B_{\mu}'(0)+\frac{1}{2}B_{\mu}''(0)r+\dots & r<R
\end{cases}\label{eq:fem-func-redef}
\end{equation}
since all matrix elements are evaluated with respect to the numerical
basis defined by \eqref{fem-func-redef}. The Taylor expansion technique
is therefore fully compatible with the variational approach pioneered
in \citeref{Lehtola2019_IJQC_25945}.

Even though \eqref{fem-func-redef} constitutes a valid definition
for the numerical basis regardless of the order of the used Taylor
expansion or the employed switch-off value $R$, it is best for accuracy
and numerical stability if the switchoff between the analytic expression
and the Taylor expansion is as smooth as possible. 

We determine the optimal switching point $R$ by maximizing the mutual
compatibility of the two definitions for the basis functions and their
derivatives at $R$. We measure the agreement with the metric
\begin{align}
\Delta(R) & =\sum_{d=0}^{2}\Delta_{d}(R)\label{eq:delta}\\
\Delta_{d}(R) & =\frac{\sum_{\mu}\left[\chi_{\mu}^{\text{analytic};(d)}(R)-\chi_{\mu}^{\text{Taylor};(d)}(R)\right]^{2}}{\sum_{\mu}\left[\chi_{\mu}^{\text{analytic};(d)}(R)\right]^{2}}\label{eq:delta-D}
\end{align}
which simultaneously considers the compatibility of the numerical
basis functions themselves as well as that of their first two derivatives.

Since the evaluation of primitive polynomials $f(r)=\sum_{n}c_{n}r^{n}$
is ill-behaved in general, and the analytical expression $\chi_{\mu}(r)=B_{\mu}(r)/r$
is only well-behaved at non-negligible $r$, we will only consider
values of $R$ in the region $0\leq R\leq r_{2}$, where $r_{2}$
is the position of the second node defining the shape functions $B_{\mu}(r)$,
the first node always being located at $r_{1}=0$ and the corresponding
basis function removed to satisfy the boundary condition $B_{\mu}(0)=0$.

\begin{figure}[H]
\subfloat[\label{fig:taylor6-lip}]{\begin{centering}
\includegraphics[width=1\linewidth]{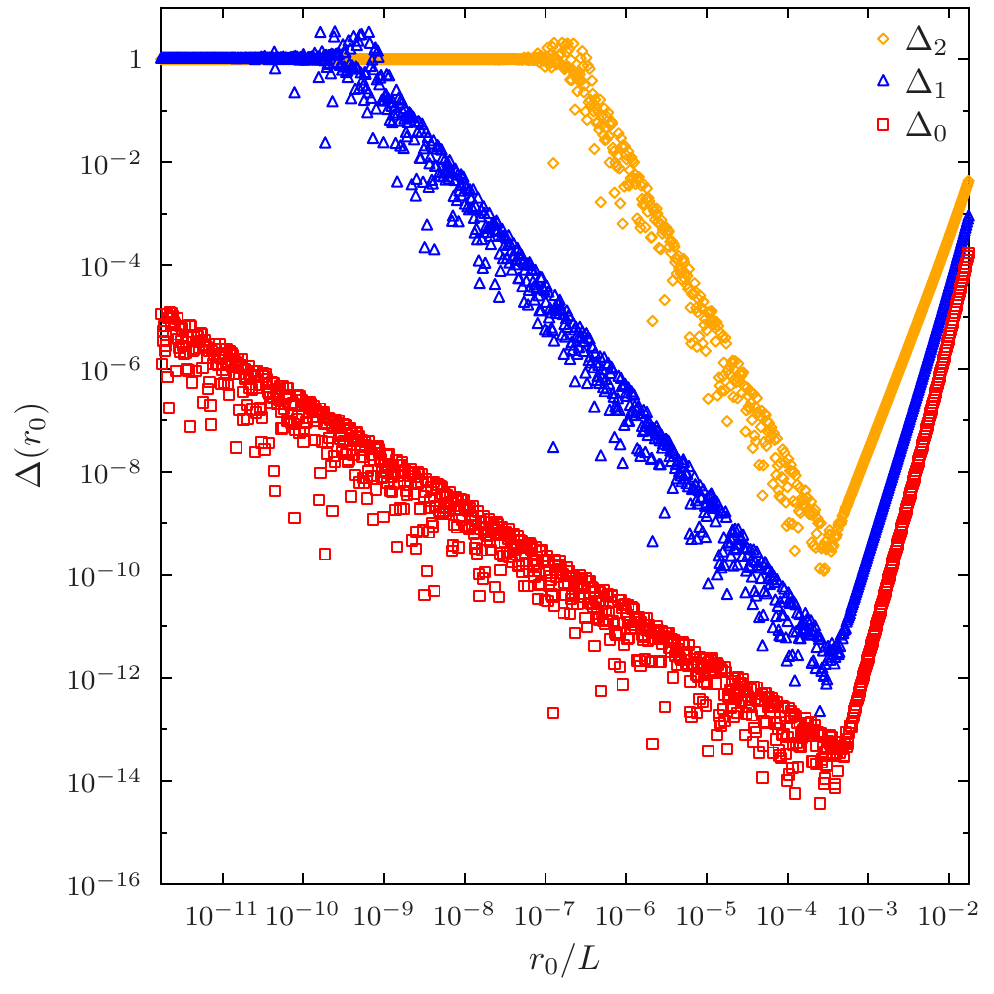}
\par\end{centering}
}

\subfloat[\label{fig:taylor6-hip}]{\begin{centering}
\includegraphics[width=1\linewidth]{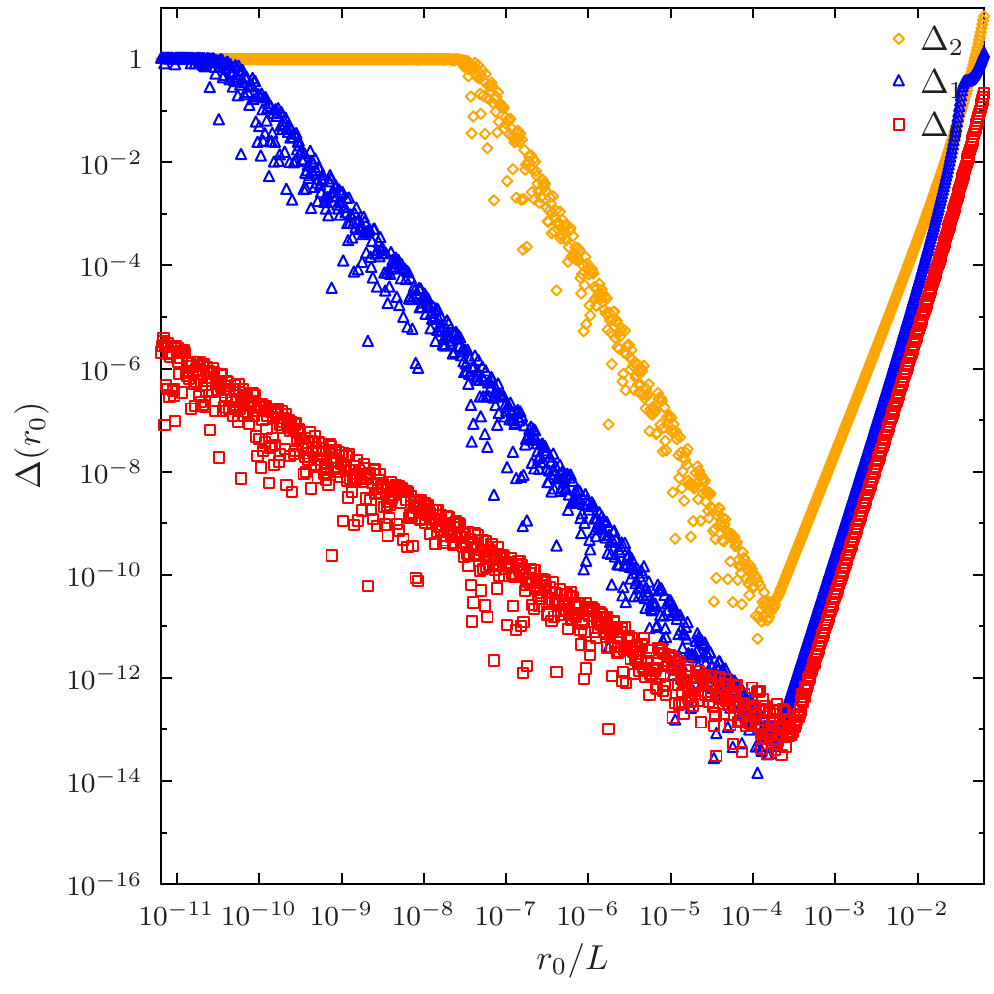}
\par\end{centering}
}

\caption{Error (\eqref{delta}) in approximate Taylor expansions of basis functions
and their derivatives (\eqref{chi-taylor}) as a function of the switchoff
radius $r_{0}$ in terms of the length $L$ of the first radial element.
\Figref{taylor6-lip} shows data for a 6$^{\text{th}}$ order Taylor
expansion in 15-node LIP basis with 5 radial elements. The first non-nuclear
node is at $r/L\approx0.017377$. \Figref{taylor6-hip} shows data
for a 6$^{\text{th}}$ order Taylor expansion in 8-node HIP basis
with 5 radial elements. The first non-nuclear node is at $r/L\approx0.064130$.
\label{fig:taylor6}}
\end{figure}

With the restriction to $0\leq R\leq r_{2}$, the fitting problem
is well-behaved. The result of a numerical experiment for a low-order
Taylor expansion is shown in \figref{taylor6}. As the analytic functions
are numerically unstable for small $R$, while the Taylor expansion
becomes inaccurate for large $R$, $\Delta(R)$ has a well-defined
minimum, but due to finite numerical precision, there is bound to
be some noise. As expected, the optimal switching value $R$ is found
to the left of the first non-nuclear node, $R<r_{2}$, and the error
$\Delta(R)$ decreases monotonically when approaching the optimal
value from the right, until it starts going back up again for $r<R$
where the analytical expression $\chi_{\mu}(r)=B_{\mu}(r)/r$ is unstable.
As the noise makes it potentially risky to pick $R$ from the global
minimum of $\Delta(R)$, we choose $R$ by proceeding downhill to
the left from $R=r_{2}$.

\Figref{taylor6} also shows that the errors increase for every derivative
when a low-order Taylor expansion is used, as the Taylor expansion
(\eqref{chi-taylor}) for the derivatives $\chi_{\mu}^{(d)}(r)$ with
$d\geq1$ have fewer and lower-order terms than the expansion for
the basis function $\chi_{\mu}(r)$. The total error in \eqref{delta}
is thereby dominated by errors in the second derivative $\chi_{\mu}''(r_{0})$
when a low-order Taylor expansion is used. 

\begin{figure}[H]
\subfloat[\label{fig:taylor-lip}]{\begin{centering}
\includegraphics[width=1\linewidth]{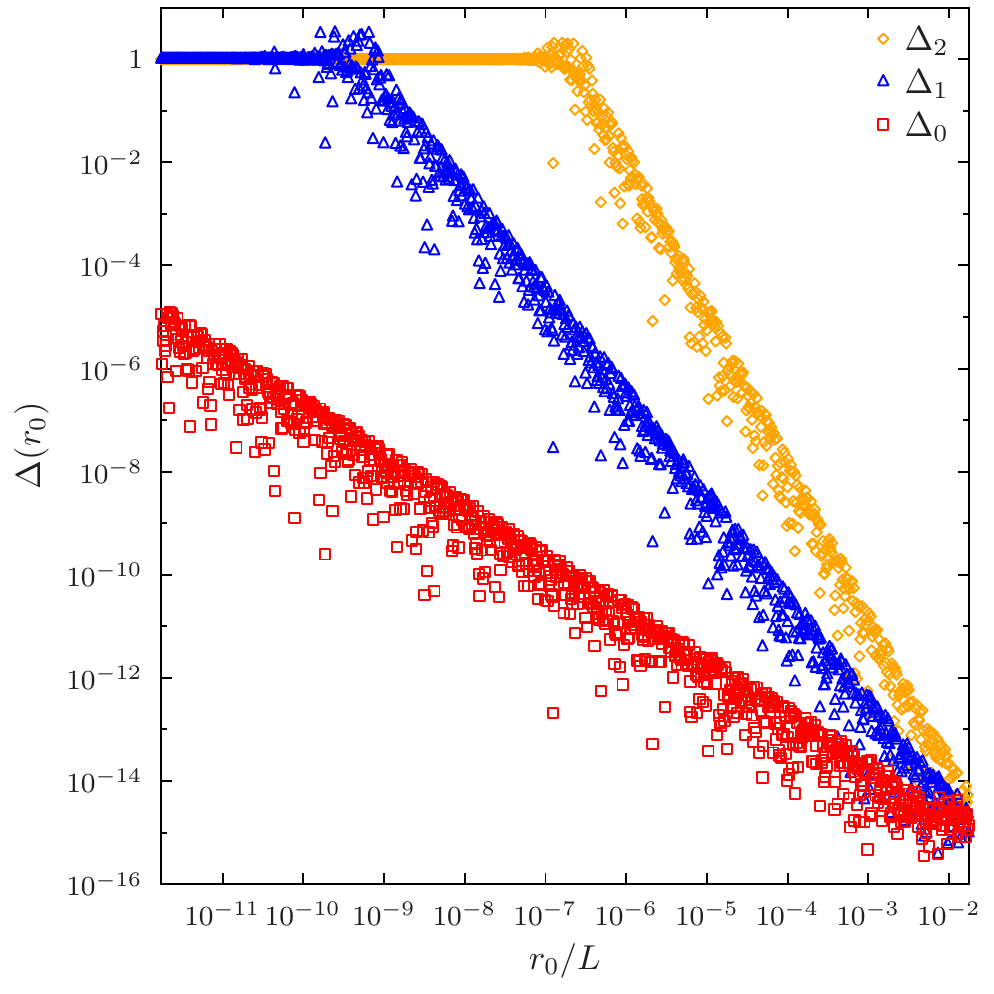}
\par\end{centering}
}

\subfloat[\label{fig:taylor-hip}]{\begin{centering}
\includegraphics[width=1\linewidth]{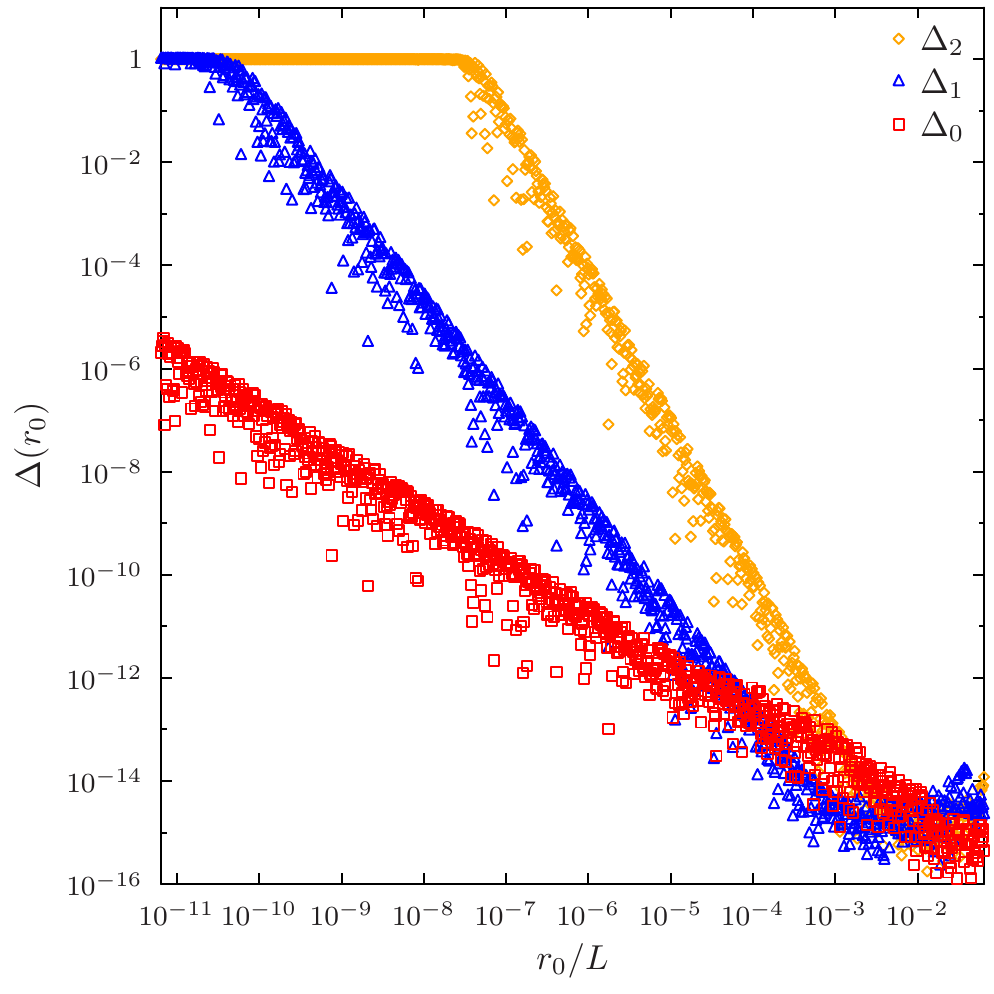}
\par\end{centering}
}

\caption{Error (\eqref{delta}) in exact Taylor expansions of basis functions
and their derivatives (\eqref{chi-taylor}) as a function of the switchoff
radius $r_{0}$ in terms of the length $L$ of the first radial element.
\Figref{taylor-lip} shows data for a 14$^{\text{th}}$ order Taylor
expansion in 15-node LIP basis with 5 radial elements. The first non-nuclear
node is at $r/L\approx0.017377$. \Figref{taylor-hip} shows data
for a 15$^{\text{th}}$ order Taylor expansion in 8-node HIP basis
with 5 radial elements. The first non-nuclear node is at $r/L\approx0.064130$.
\label{fig:taylor}}
\end{figure}

However, when the order of the Taylor series is increased, we observe
that the optimal value of $R$ moves to the right, and the corresponding
minimal value $\Delta(R)$ goes down. When a Taylor expansion of an
order matching that of the shape function basis $B_{\mu}(r)$ is used,
the optimal $\Delta(R)$ is practically zero, as shown in \figref{taylor},
and the switchoff value moves all the way to the right, becoming $R=r_{2}$. 

Since the analytic derivatives of the shape functions described in
\subsecref{Basis-Polynomials} are easy to generate to arbitrarily
high orders (at present the code supports up to $20^{\text{th}}$
order Taylor expansions), in the following we employ such full-length
Taylor expansions in all calculations, which is also the new default
in \HelFEM{}.

\subsection{Supremacy of HIPs over LIPs \label{subsec:suphip}}

So far the study has used the pre-established finite element grid
from \citeref{Lehtola2019_IJQC_25945}. However, since the grid was
optimized in \citeref{Lehtola2019_IJQC_25945} for LIPs and noble
gas atoms at the HF level of theory, it is worthwhile to check whether
the same grid also works well with HIPs and density functionals of
various rungs. It is interesting to note that an $n^{\text{th}}$
order HIP calculation (LIPs correspond to $n=0$, first-order HIPs
to $n=1$) has 
\begin{equation}
N_{\text{bf}}^{\text{HIP}n}=(n+1)N_{\text{elem}}(N_{\text{nodes}}-1)-1\label{eq:nhip}
\end{equation}
basis functions, if the derivatives are set to zero at the practical
infinity.\bibnote{There are altogether $(n+1)N_{\text{nodes}}N_{\text{elem}}$ shape functions in the elements, out of which $(n+1)(N_{\text{elem}}-1)$ are at element boundaries and are therefore overlaid, and the first function value (not the derivatives!) is removed for the boundary condition at the nucleus, and the last function value and all $n$ derivatives are removed for the boundary conditions at the practical infinity. }
This means that apples-to-apples comparisons of numerical basis sets
for various orders is possible by choosing values for the order $n$
and number of nodes $N_{\text{nodes}}$ whose product yields a constant
$(n+1)(N_{\text{nodes}}-1)$, such as $N_{\text{nodes}}^{\text{LIP}}=2N_{\text{nodes}}^{\text{HIP}}-1$
for comparisons of LIPs with 1st order HIPs.

\begin{figure*}
\subfloat[HF \label{fig:Truncation-error-hf}]{\begin{centering}
\includegraphics[width=0.5\linewidth]{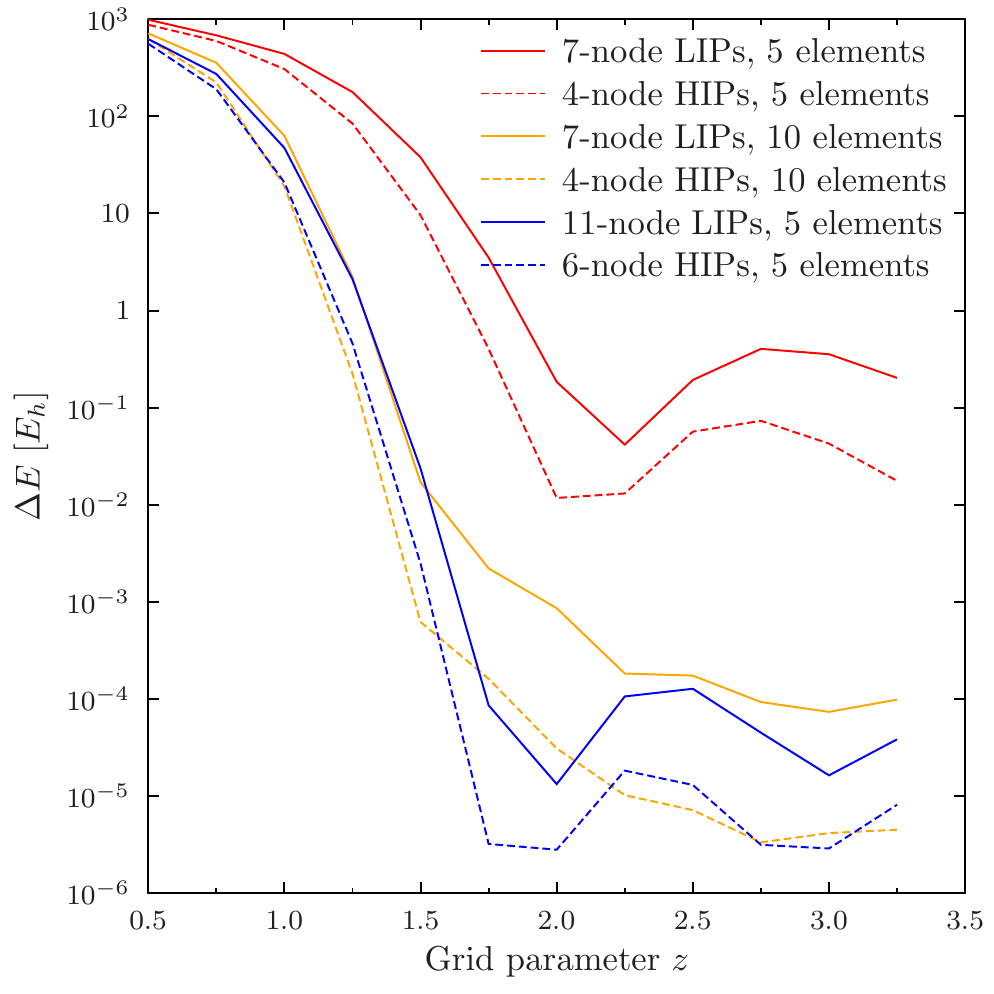}
\par\end{centering}
}\subfloat[PW92 \label{fig:Truncation-error-pw92}]{\begin{centering}
\includegraphics[width=0.5\linewidth]{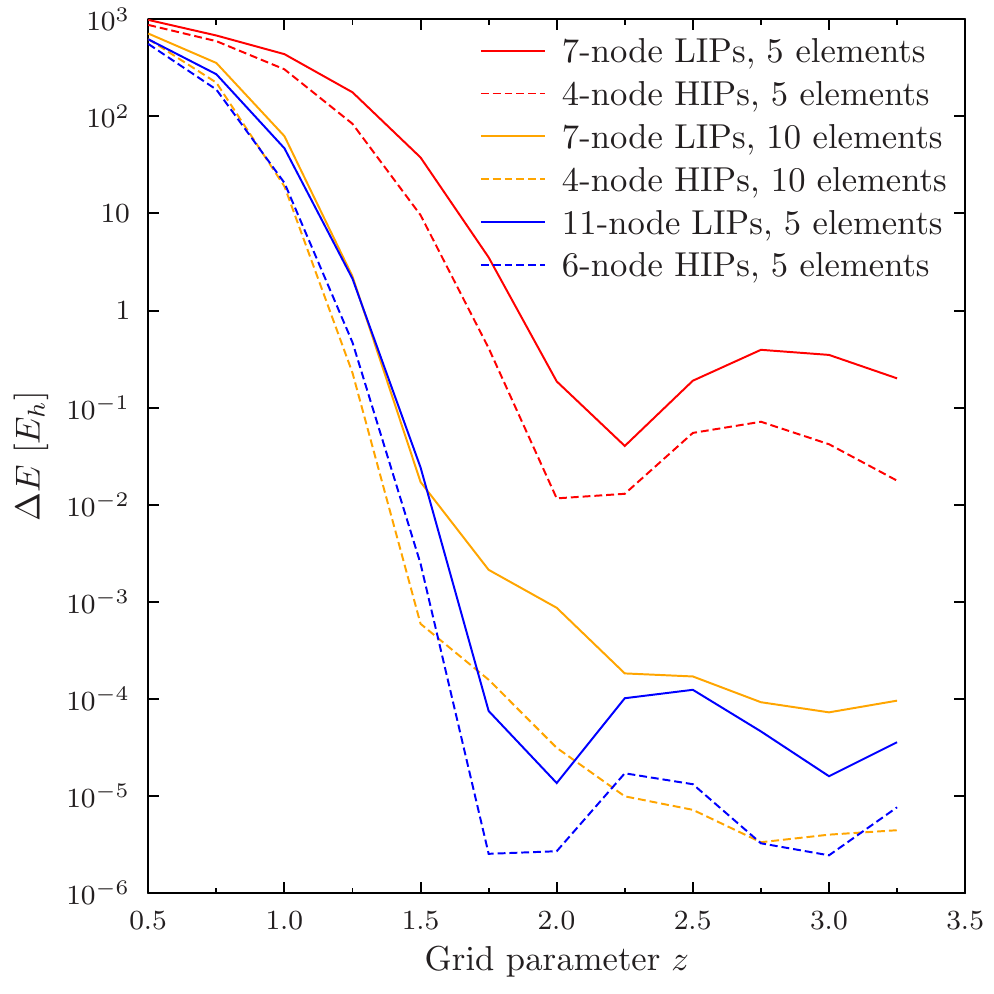}
\par\end{centering}
}

\subfloat[PBE \label{fig:Truncation-error-pbe}]{\begin{centering}
\includegraphics[width=0.5\linewidth]{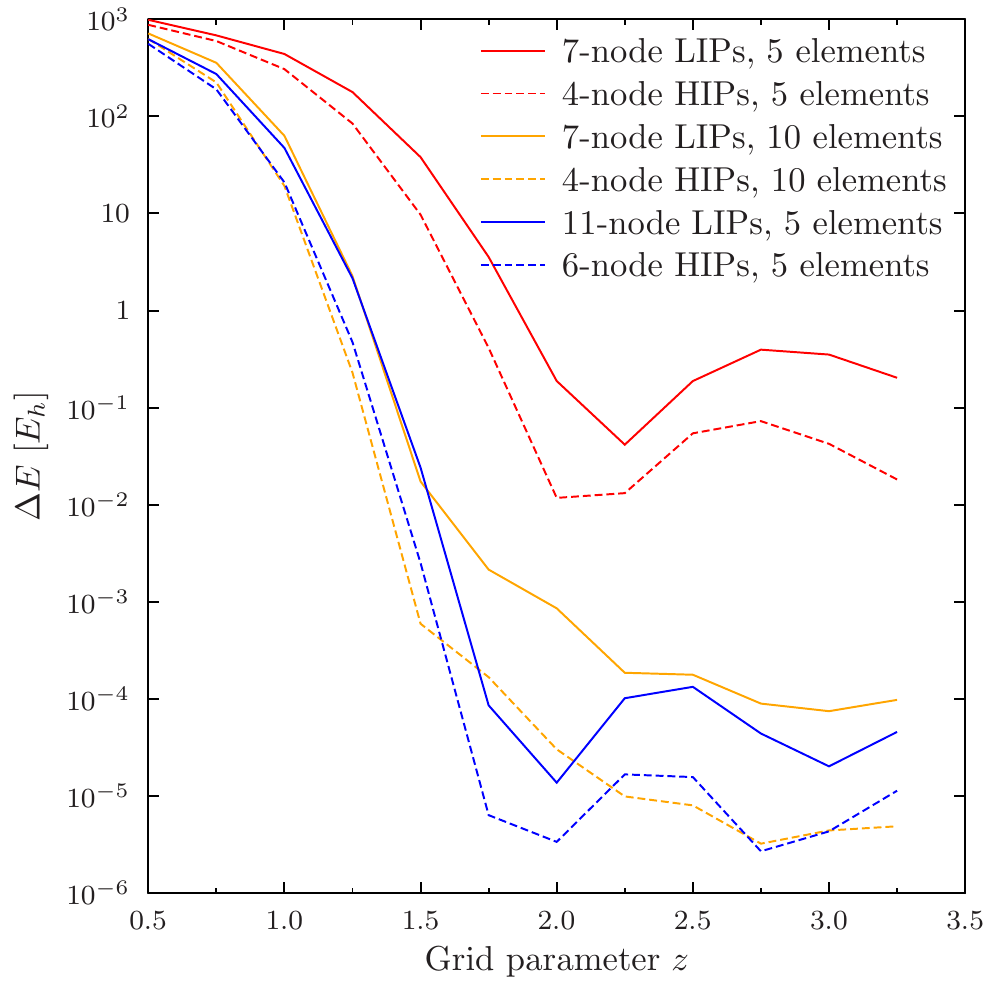}
\par\end{centering}
}\subfloat[TASKCC \label{fig:Truncation-error-taskcc}]{\begin{centering}
\includegraphics[width=0.5\linewidth]{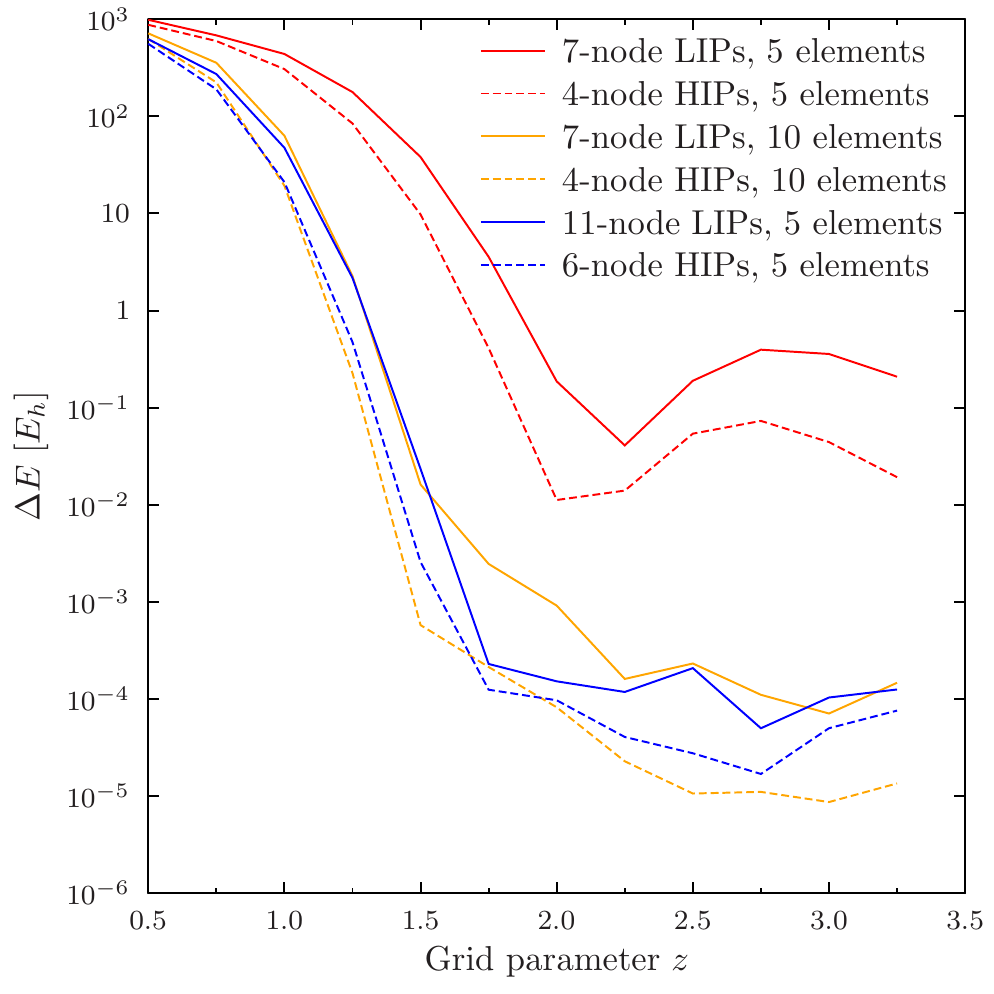}
\par\end{centering}
}

\caption{Truncation error $\Delta E(z)=E(z)-E^{\text{CBS}}$ in total energy
of Zn as a function of the finite element grid parameter $z$ (\eqref{expgrid})
for (\ref{fig:Truncation-error-hf}) HF, (\ref{fig:Truncation-error-pw92})
PW92, (\ref{fig:Truncation-error-pbe}) PBE, and (\ref{fig:Truncation-error-taskcc})
TASKCC. Note logarithmic $y$ axis. \label{fig:Truncation-error}}
\end{figure*}

An illustration of the superior accuracy of the HIP basis over the
LIP basis is shown in \figref{Truncation-error}, which studies optimal
choices for the radial finite element grid to compute the total energy
of Zn for such compatible choices for LIPs ($n=0$) and the analytic
first-order HIPs ($n=1$). In addition to HF, which was used in \citeref{Lehtola2019_IJQC_25945}
to determine the recommended value $z=2$, \figref{Truncation-error}
also considers the \citeyear{Perdew1992_PRB_13244} LDA correlation
functional by \citet{Perdew1992_PRB_13244} used in combination with
LDA exchange\citep{Bloch1929_ZfuP_545,Dirac1930_MPCPS_376} (PW92),
the \citeyear{Perdew1996_PRL_3865} GGA exchange-correlation functional
by \citet{Perdew1996_PRL_3865} (PBE), and the \citeyear{Aschebrock2019_PRR_33082}
meta-GGA exchange functional of \citet{Aschebrock2019_PRR_33082}
in combination with the meta-GGA correlation functional of \citet{Schmidt2014_JCP_18}
(TASKCC) as suggested by \citet{Lebeda2022_PRR_23061}; all density
functionals are evaluated in \HelFEM{} with \Libxc{}.\citep{Lehtola2018_S_1}

\Figref{Truncation-error} again demonstrates that the choice for
the grid is extremely important, as the quality of the resulting wave
function is entirely dependent on a suitable distribution of the degrees
of freedom. Non-uniform grids afford similarly good results with LIPs
and HIPs. \Figref{Truncation-error}  also supports the general agreement
based on Gaussian-basis calculations\citep{Jensen2002_JCP_7372} that
the basis set requirements of HF and DFT calculations are similar:
the truncation errors for HF, PW92, PBE and TASKCC are similar in
behavior and magnitude.

As \figref{Truncation-error} shows, improvements of roughly an order
of magnitude are possible when switching over from a LIP basis to
a HIP basis with the same total number of basis functions. However,
this only applies when the basis sets are limited to a low numerical
order: when a large enough numerical basis set is used, either basis
converges to the same total energy. 

As was discussed in \citeref{Lehtola2019_IJQC_25945}, the convergence
of the SCF energy to the CBS limit is extremely rapid (superexponential)
with the number of basis functions, when the number of nodes is also
increased; note that this roughly corresponds to $hp$-adaptive FEM
where both the discretization ($h$) and the order of the polynomial
basis (\textbf{$p$}) is changed, even though our grids are merely
empirically optimal. By default, \HelFEM{} employs numerical basis
functions of a high order, and the CBS limit can be routinely reached
by simply adding more radial elements until the total energy does
not change any more.

\begin{figure*}
\subfloat[LIP basis, polynomial order $=N_{\text{nodes}}-1$. \label{fig:Xe-LIP}]{\begin{centering}
\includegraphics[width=0.5\linewidth]{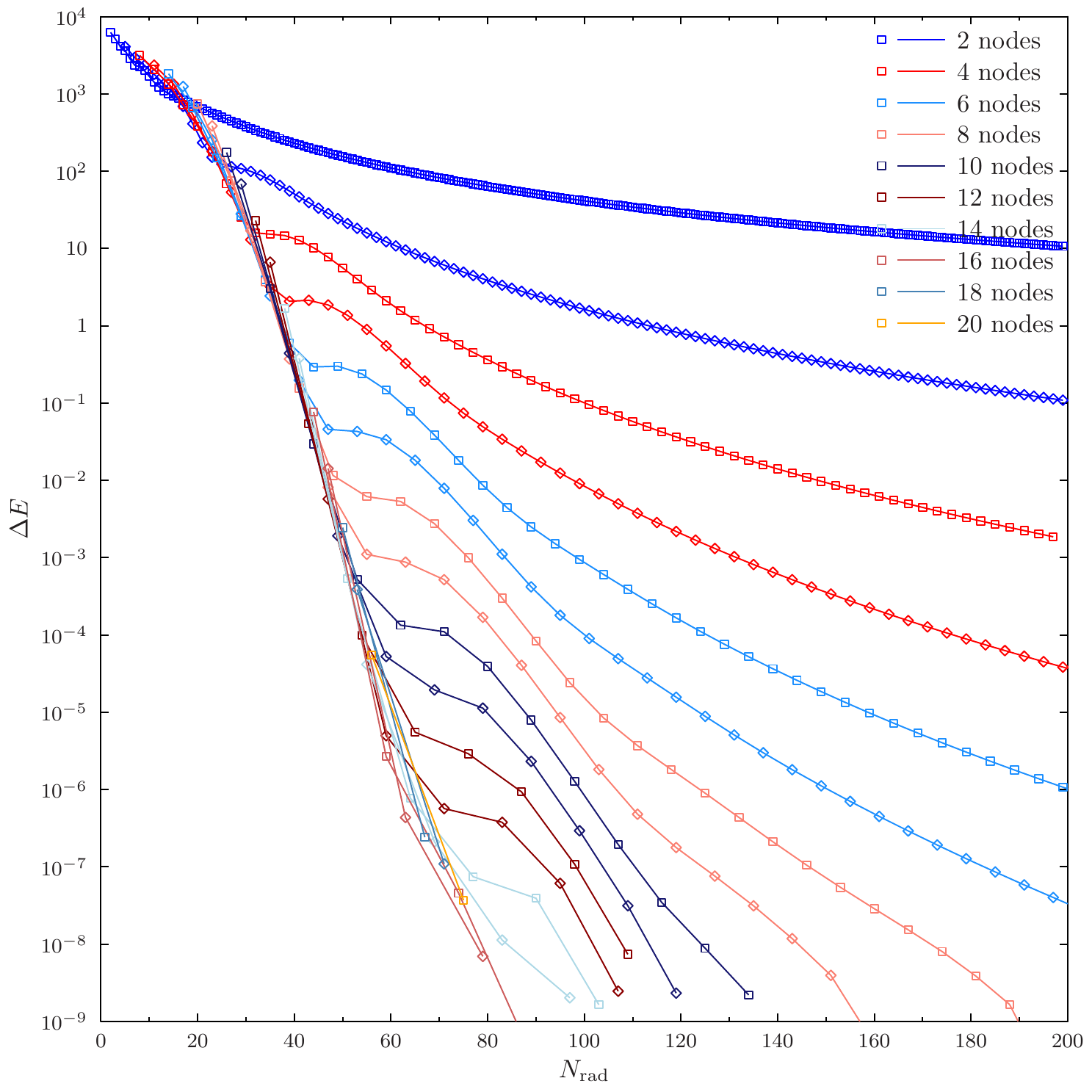}
\par\end{centering}
}\subfloat[Analytic first-order HIP basis, polynomial order $=2N_{\text{nodes}}-1$.
\label{fig:Xe-HIP}]{\begin{centering}
\includegraphics[width=0.5\linewidth]{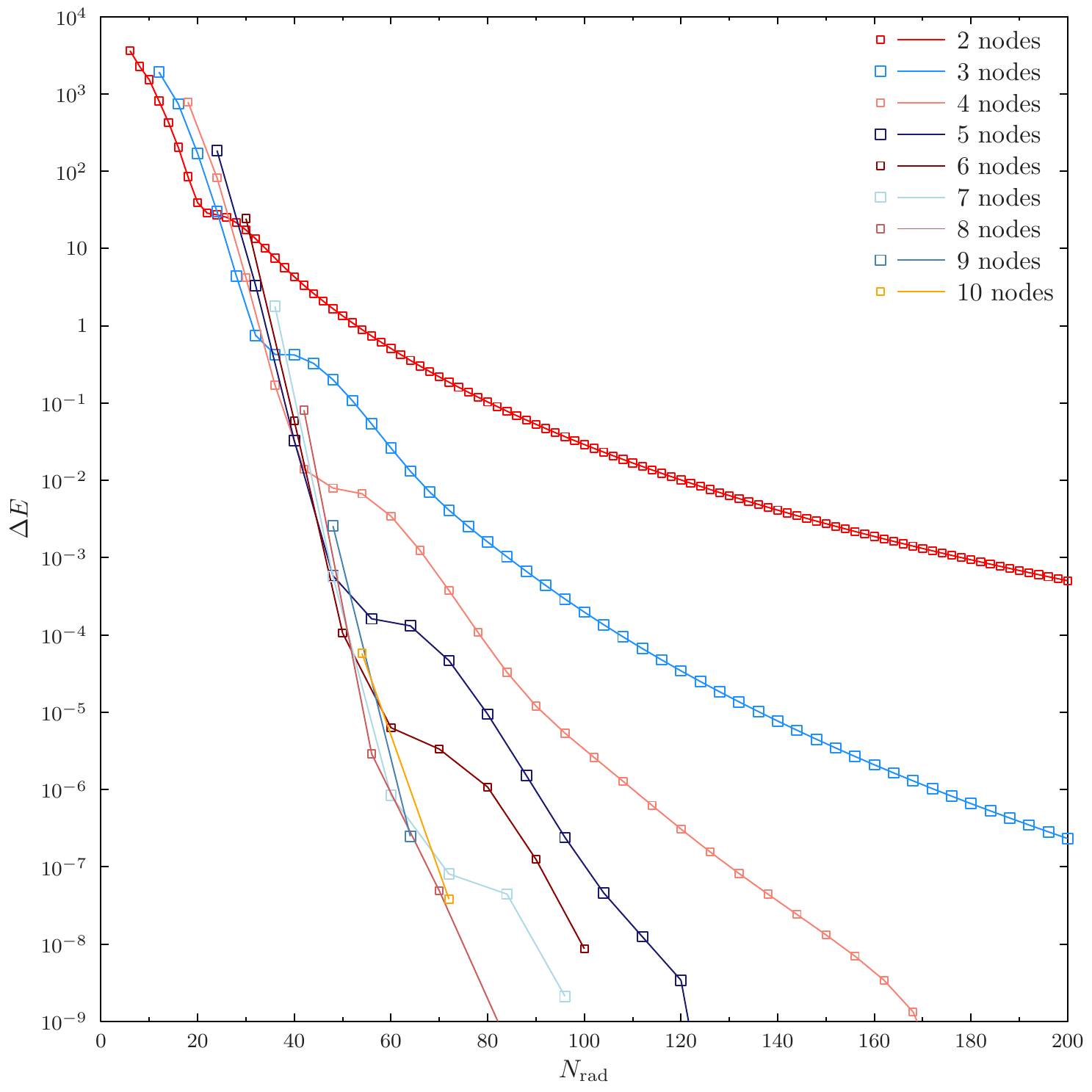}
\par\end{centering}
}

\subfloat[Numerical second-order HIP basis, polynomial order $=3N_{\text{nodes}}-1$.
\label{fig:Xe-n2HIP}]{\begin{centering}
\includegraphics[width=0.5\linewidth]{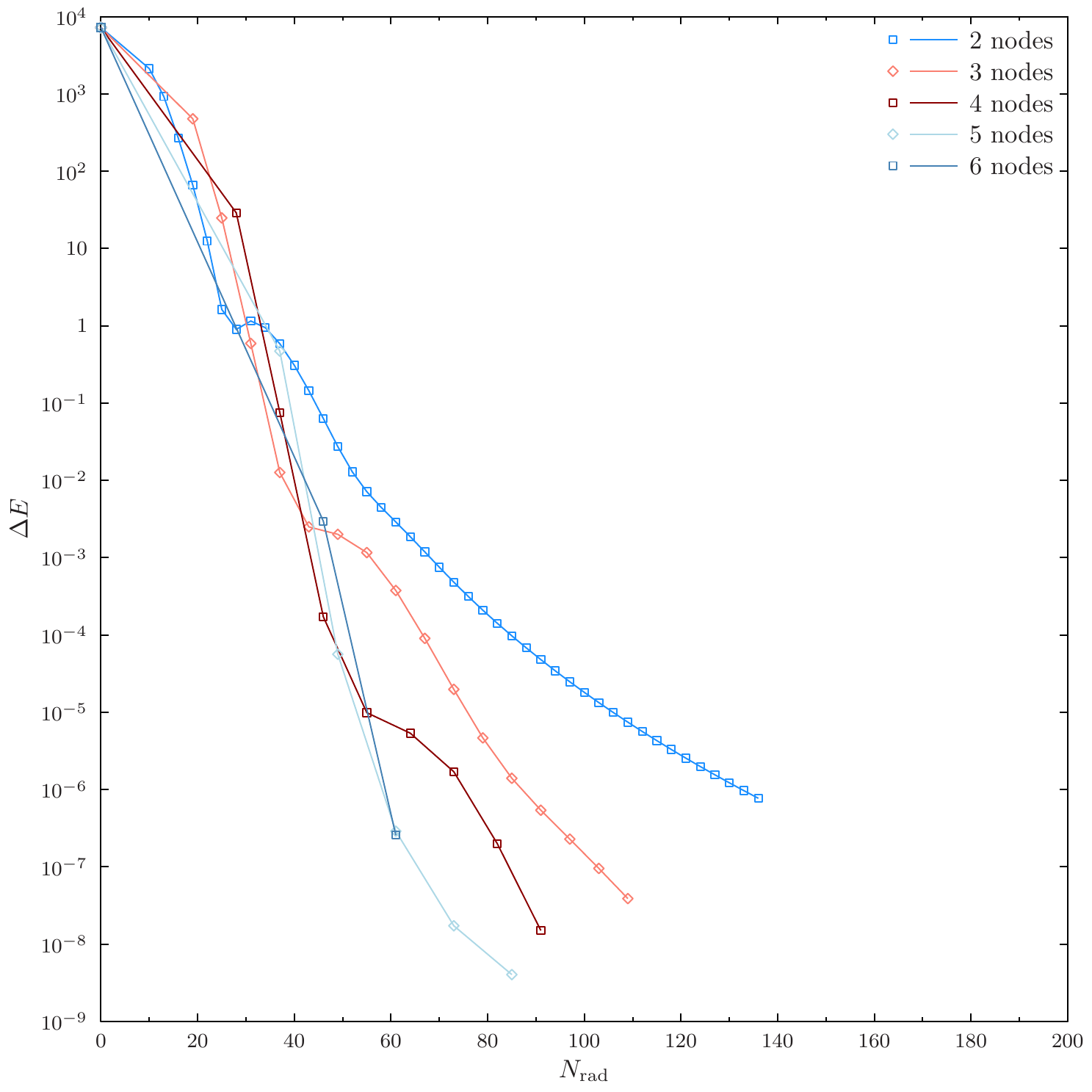}
\par\end{centering}
}

\caption{Truncation error $\Delta E=E-E^{\text{CBS}}$ in total Hartree--Fock
energy of the Xe atom for (a) LIP, (b) analytical first-order HIP
and (c) numerical second-order HIP basis sets. Note logarithmic $y$
axis. The CBS limit energy is $E^{\text{CBS}}=-7232.13836387$ as
determined by \citet{Saito2009_ADNDT_836}. In \figref{Xe-LIP}, plots
for even numbers of nodes are shown with solid lines marked with squares,
as shown in the legend, while plots for odd numbers of nodes are shown
with solid lines marked with diamonds, the basis having one more node
than the even-numbered calculation of the same color. The colors and
markers in \figref{Xe-HIP, Xe-n2HIP} are chosen such that the order
of the polynomial basis matches in \figref{Xe-LIP, Xe-HIP, Xe-n2HIP}.
\label{fig:HF-CBS}}
\end{figure*}

In pratice, either basis (LIP or HIP) can be used to converge the
total energy to the CBS limit, which is the whole point of employing
fully numerical methods. At the end, it should not matter which numerical
method is used to obtain results, only that the results are fully
converged to the CBS limit. Following \citeref{Lehtola2019_IJQC_25945},
the power of high-order numerical schemes is demonstrated by the truncation
errors plotted in \figref{HF-CBS} for the studied LIP and HIP methods. 

We observe that the accuracy obtained with LIPs, analytical first-order
HIPs, as well as numerical second-order HIPs is highly affected by
the employed polynomial order, as is demonstrated by the rapid decrease
of the truncation error at fixed number of radial basis functions
$N_{\text{rad}}$ with increasing polynomial order of the basis; this
finding was one of the main results of \citeref{Lehtola2019_IJQC_25945},
but that study was limited to LIPs.

We also see great similarities between \figref{Xe-LIP,Xe-HIP} as
well as \figref{Xe-LIP,Xe-n2HIP}: the plots for the same polynomial
order with LIPs and HIPs have almost the same shape, the HIP calculation
just having fewer basis functions in total, since more functions are
overlaid as discussed above.

\subsection{HIPs and meta-GGA functionals \label{subsec:metaGGA}}

Next, we will look deeper into how the numerical basis set affects
the calculations with meta-GGA functionals for the exchange-correlation
(xc) energy, which depend on the spin-$\sigma$ local kinetic energy
density 
\begin{equation}
\tau_{\sigma}(\boldsymbol{r})=\frac{1}{2}\sum_{i\text{ occupied}}\left|\nabla\psi_{i\sigma}(\boldsymbol{r})\right|^{2}\label{eq:tau}
\end{equation}
and/or the Laplacian of the spin-$\sigma$ electron density $\nabla^{2}n_{\sigma}$
as %\begin{widetext}

\begin{align}
E_{\text{xc}}= & \int n(\boldsymbol{r})\epsilon_{\text{xc}}(\{n_{\sigma}(\boldsymbol{r})\},\{\gamma_{\sigma\sigma'}(\boldsymbol{r})\},\nonumber \\
 & \{\tau_{\sigma}(\boldsymbol{r})\},\{\nabla^{2}n_{\sigma}(\boldsymbol{r})\}){\rm d}^{3}r,\label{eq:Exc}
\end{align}
%\end{widetext} where $\gamma_{\sigma\sigma'}(\boldsymbol{r})=\nabla n_{\sigma}(\boldsymbol{r})\cdot\nabla n_{\sigma'}(\boldsymbol{r})$
is the reduced gradient and $\epsilon_{\text{xc}}$ is the used density
functional approximation (DFA) that models the energy density per
particle.

\subsubsection{Exact Results for the Hydrogen Atom \label{subsec:exactH}}

A simple test is offered by the hydrogen atom, whose exact non-relativistic
Born--Oppenheimer ground state is $\psi(\boldsymbol{r})=\pi^{-1/2}e^{-r}$.
As the wave function is spherically symmetric, we will examine $\tau(r)$
and $[\nabla^{2}n](r)$ which represent $\tau(\boldsymbol{r})$ and
$\nabla^{2}n(\boldsymbol{r})$ integrated over all angles: 
\begin{equation}
\tau(r)=2e^{-2r}\label{eq:h-tau}
\end{equation}
 and
\begin{equation}
\nabla^{2}n(r)=16\left(1-\frac{1}{r}\right)e^{-2r}.\label{eq:h-lapl}
\end{equation}
Note that \eqref{h-lapl} is negative for $r<1$, diverging to $-\infty$
for $r\to0$, and is positive for $r>1$. 

Before proceeding to the CBS limit, it is useful to study results
that are not fully converged with respect to the basis set. We use
HF for this demonstration, as it is exact for the H atom. We employ
5 radial elements and study a 7-node LIP basis (29 radial basis functions,
$E=-0.4999993E_{h}$), a 4-node analytic first-order HIP basis set
(29 radial basis functions, $E=-0.4999999E_{h}$), and a 3-node numerical
second-order HIP basis set (29 radial basis functions, $E=-0.5000000E_{h}$)
that is expressed in terms of an underlying 9-node LIP basis. 

\begin{figure}[H]
\begin{centering}
\subfloat[$\tau$ \label{fig:H-tau-small}]{\begin{centering}
\includegraphics[width=1\linewidth]{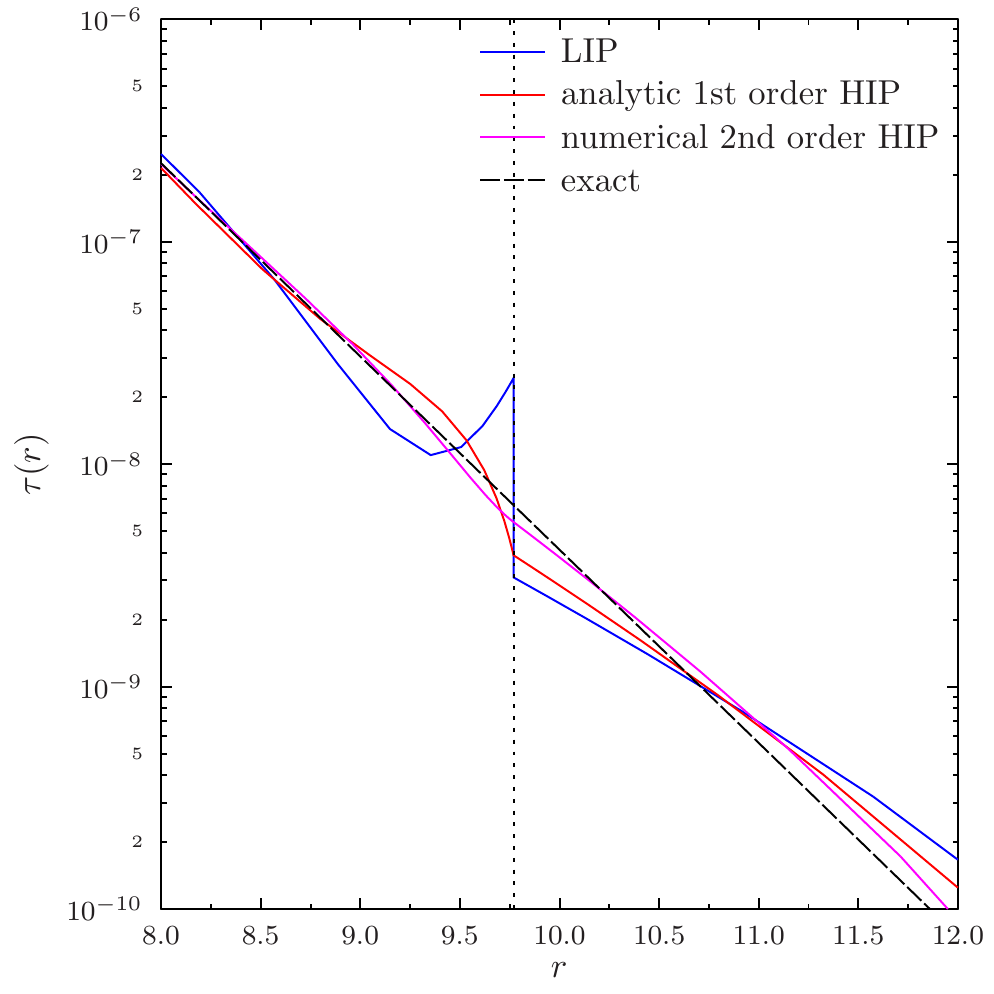}
\par\end{centering}
}
\par\end{centering}
\begin{centering}
\subfloat[$\nabla^{2}n$ \label{fig:H-lapl-small}]{\begin{centering}
\includegraphics[width=1\linewidth]{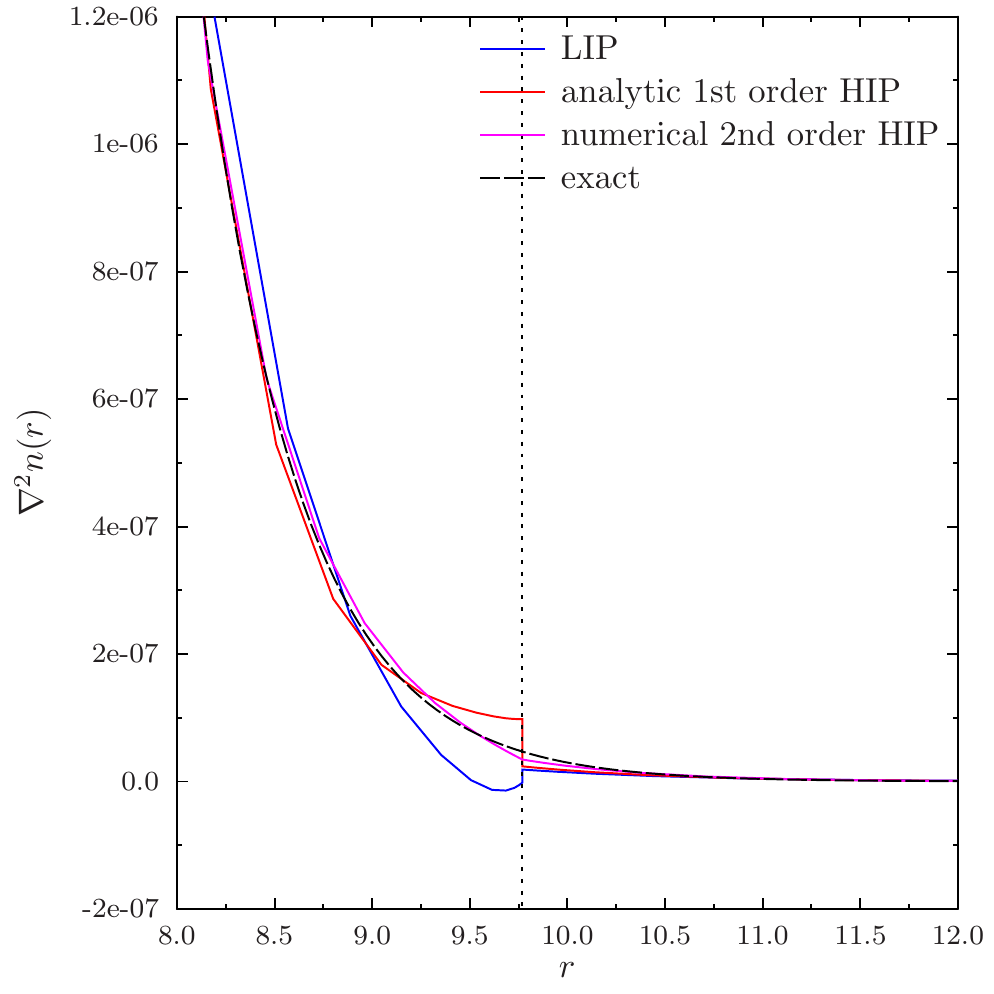}
\par\end{centering}
}
\par\end{centering}
\caption{Closeup of discontinuities in $\tau$ and $\nabla^{2}n$ of the ground
state of the hydrogen atom in calculations not converged to the CBS
limit. The exact values are computed with \eqref{h-tau, h-lapl},
respectively, and the boundary between the fourth and fifth radial
elements is shown with the vertical dotted line. \label{fig:H-lapltau-small}}
\end{figure}

We observe that there are discontinuities in $\tau$ and $\nabla^{2}n$
between the fourth and fifth radial elements, shown in \figref{H-lapltau-small},
even though all the numerical energies are very close to the exact
value $E=-0.5E_{h}$. As can be seen from \figref{H-tau-small}, the
LIP basis shows a step discontinuity already in $\tau$, which is
also accompanied by a step discontinuity in $\nabla^{2}n$ as seen
from \figref{H-lapl-small}. The first-order analytic HIP curve shows
a cusp in $\tau$, that is, a discontinuity of the first derivative
of $\tau$, while $\nabla^{2}n$ is still stepwise discontinuous.
The numerical second-order HIP basis, on the other hand, produces
a smooth $\tau$ but still has a noticeable cusp in $\nabla^{2}n$
at the element boundary.

However, these issues go away when the size of the numerical basis
set is increased. In the rest of this section, we will consider 15-node
LIPs, 8-node analytic first-order HIPs, and 5-node numerical second-order
HIPs that are expressed in terms of a 15-node LIP basis, all with
five radial elements. Each of these basis sets reproduces a HF total
energy of $-0.5000000E_{h}$. 

\begin{figure}[H]
\subfloat[$\tau$ \label{fig:H-tau}]{\begin{centering}
\includegraphics[width=1\linewidth]{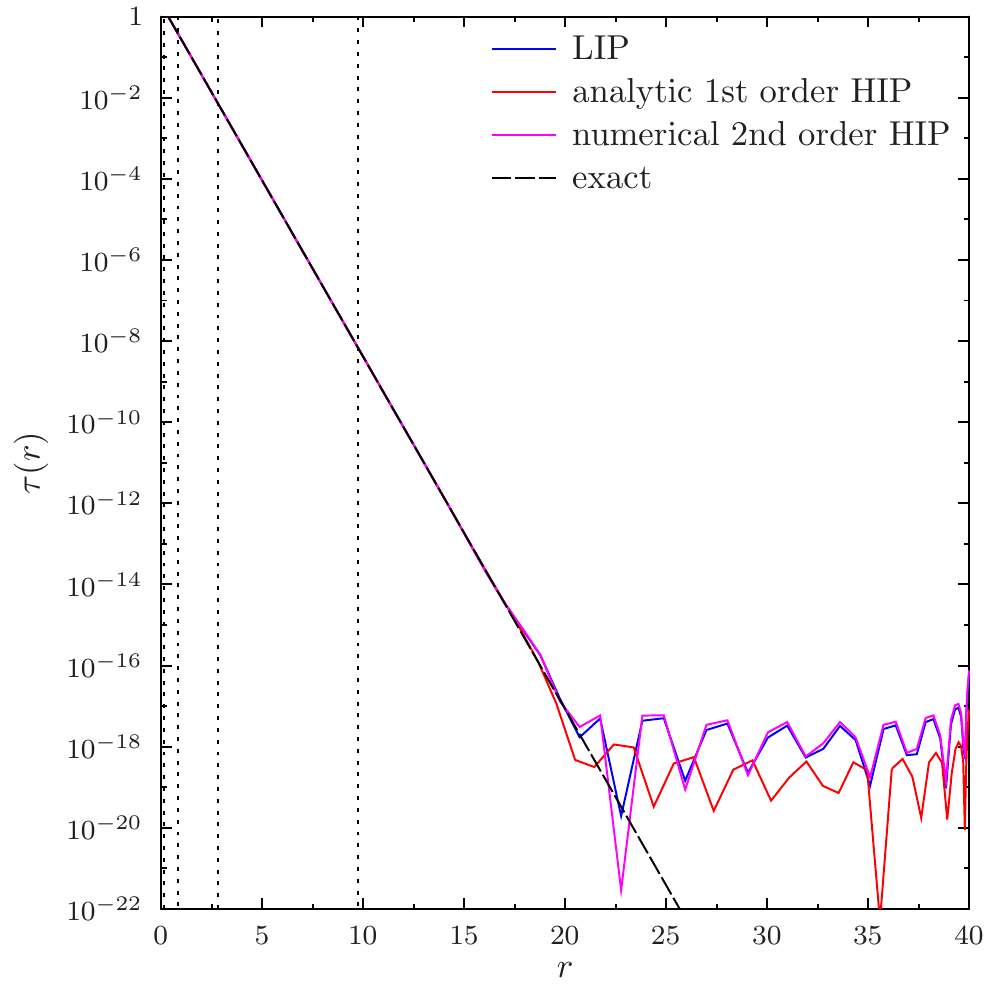}
\par\end{centering}
}

\subfloat[$\nabla^{2}n$ \label{fig:H-lapl}]{\begin{centering}
\includegraphics[width=1\linewidth]{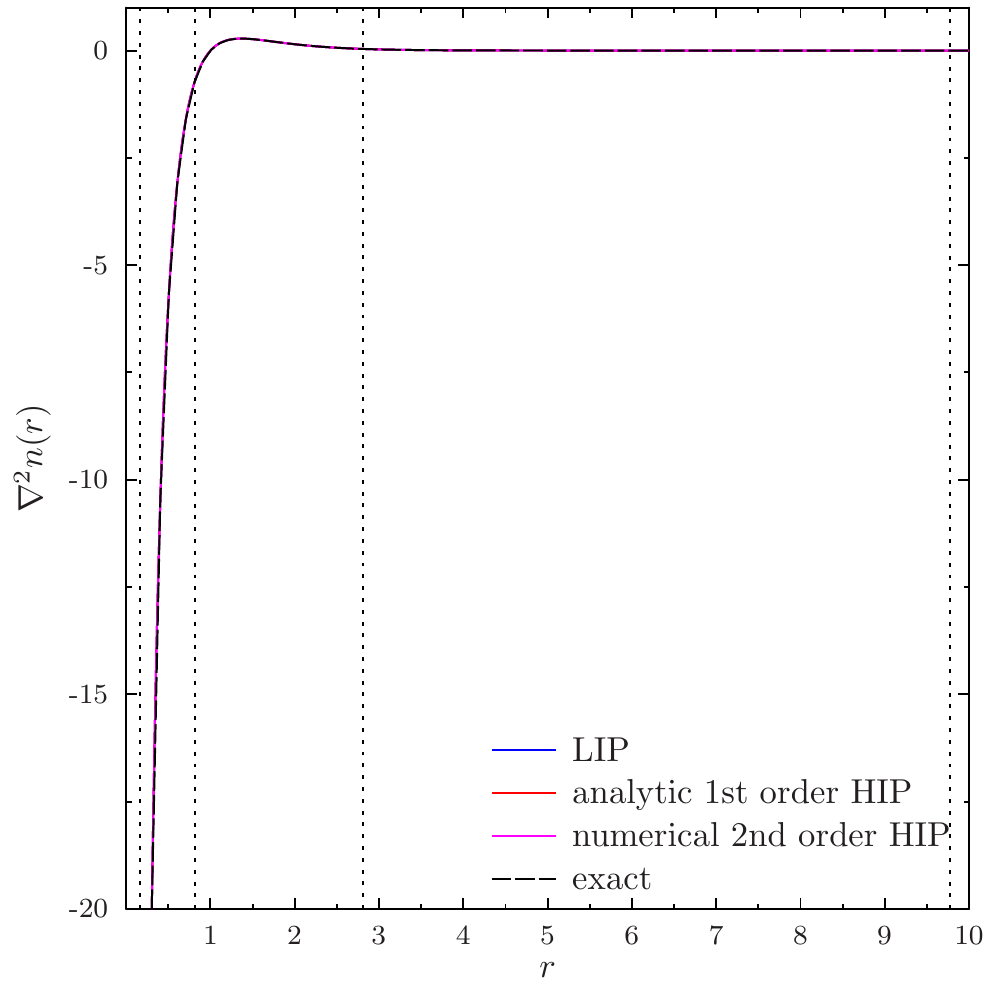}
\par\end{centering}
}

\caption{$\tau$ and $\nabla^{2}n$ of the ground state of the hydrogen atom,
computed with HF with either a 15-node LIP basis, an 8-node analytic
first-order HIP basis set, or a 5-node numerical second-order HIP
basis set. Five radial elements were used in all calculations. The
exact values are computed with \eqref{h-tau, h-lapl}, respectively,
and the boundaries between the radial elements are shown with the
vertical dotted lines. \label{fig:H-lapltau}}
\end{figure}

We have previously argued that the energy minimization involved in
the variational theorem of self-consistent field theory should ensure
a smooth wave function, even if the underlying numerical basis set
does not guarantee explicit continuity.\citep{Lehtola2019_IJQC_25945}
This indeed does appear to be the case: we see that both $\tau$ and
$\nabla^{2}n$ are successfully reproduced for the hydrogen atom by
the larger numerical basis sets without problems, as shown in \figref{H-tau}
for $\tau$ and \figref{H-lapl} for $\nabla^{2}n$.

\subsubsection{Self-consistent Calculations \label{subsec:taufun}}

We have now established that all three numerical basis sets are able
to reproduce $\tau$ and $\nabla^{2}n$ in HF calculations. We note
that we have recently analyzed a thorough selection of density functionals
in \citeref{Lehtola2022_JCP_174114} and found <that many recent meta-GGAs
are numerically ill-behaved at fixed density. We wish to continue
this work here by studying how the smoothness of the numerical basis
set used to represent the orbitals can affect the behavior of the
density functional when the density is relaxed. The form of the numerical
basis may be important in calculations with meta-GGA functionals,
since the functionals depend explicitly on $\tau$ and/or $\nabla^{2}n$
through \eqref{Exc}. 

We note that $\tau$-dependent functionals form an overwhelming majority
of available meta-GGAs,\citep{Lehtola2018_S_1} and that many of the
few available Laplacian dependent functionals were found ill-behaved
in \citeref{Lehtola2022_JCP_174114}. For this reason, we will only
discuss $\tau$-dependent functionals in this work. We have previously
described successful calculations with $\tau$-dependent meta-GGA
functionals with LIP basis sets,\citep{Lehtola2019_IJQC_25945,Lehtola2019_IJQC_25944,Lehtola2021_JCTC_943}
and we will critically study such calculations here. 

We begin by the general note that we have observed that calculations
with meta-GGAs often fail when the numerical basis is not large enough
for CBS limit quality, such as when employing only one or two radial
elements instead of the five elements employed to obtain these results;
similar observations were made with both LIPs and the first- and second-order
HIPs. Such convergence problems can be understood through the discussion
on the exact ground state in \subsecref{suphip}: a basis set that
is not sufficiently flexible can produce spurious cusps or oscillations
in the optimal $\tau$ (example in \figref{H-lapltau-small}), whereas
the optimal $\tau$ in a more complete basis is likely smoother and
thereby easier to find. Meta-GGA calculations should therefore use
numerical basis sets that can reproduce the CBS limit energy for HF
at the very least.

Our selection of meta-GGA functionals starts out with the TPSS functional
of \citet{Tao2003_PRL_146401} and TASKCC\citep{Aschebrock2019_PRR_33082,Schmidt2014_JCP_18}.
The r$^{2}$SCAN functional\citep{Furness2020_JPCL_8208,Furness2020_JPCL_9248}
has been found ill-behaved in a recent fully numerical study;\citep{Holzwarth2022_PRB_125144}
we also found r$^{2}$SCAN to be ill-behaved at fixed electron density
in \citeref{Lehtola2022_JCP_174114}, and choose to add it to our
present study. For completeness, we also study a number of semiempirical
meta-GGA functionals, which might present other types of issues. We
limit ourselves to functionals that did not appear problematic at
fixed electron densities.\citep{Lehtola2022_JCP_174114} 

Our selection includes two major families of modern semiempirical
density functionals. First, we have the Minnesota family composed
of the M05,\citep{Zhao2005_JCP_161103} M05-2X,\citep{Zhao2006_JCTC_364}
M06,\citep{Zhao2008_TCA_215} M06-2X,\citep{Zhao2008_TCA_215} M11,\citep{Peverati2011_JPCL_2810}
MN12-SX,\citep{Peverati2012_PCCP_16187} and MN15\citep{Yu2016_CS_5032}
hybrid functionals as well as their local versions M06-L,\citep{Zhao2006_JCP_194101}
M11-L,\citep{Peverati2012_JPCL_117} MN12-L\citep{Peverati2012_PCCP_13171}
and MN15-L,\citep{Yu2016_JCTC_1280} respectively, accompanied by
the most recent additions: the \citeyearpar{Wang2017_PNASUSA_8487}
revM06-L functional,\citep{Wang2017_PNASUSA_8487} the \citeyearpar{Wang2018_PNASUSA_10257}
revM06 functional,\citep{Wang2018_PNASUSA_10257} \citeyearpar{Verma2019_JPCA_2966}
revM11\citep{Verma2019_JPCA_2966} functional, and the \citeyearpar{Wang2020_PNASUSA_2294}
M06-SX functional.\citep{Wang2020_PNASUSA_2294} Second, the Berkeley
family is formed by the $\omega$B97X-V\citep{Mardirossian2014_PCCP_9904}
hybrid GGA, the B97M-V\citep{Mardirossian2015_JCP_74111} meta-GGA
and $\omega$B97M-V\citep{Mardirossian2016_JCP_214110} hybrid meta-GGA
functionals by Mardirossian and Head-Gordon. As it is well-known that
the Vydrov--van Voorhis non-local correlation\citep{Vydrov2010_JCP_244103}
(VV10) does not change the electron density significantly,\citep{Najibi2018_JCTC_5725}
we will only consider $\omega$B97X-V, B97M-V, and $\omega$B97M-V
without the non-local correlation part, which we denote as $\omega$B97X-noV,
B97M-noV, and $\omega$B97M-noV, respectively.

The comparison of the Minnesota and Berkeley meta-GGA functionals
is interesting, as the functionals utilize similar ingredients: $\tau$
dependence is expressed in terms of the normalized variable $-1\le w_{\sigma}\leq1$
given by
\[
t_{\sigma}=\frac{\tau_{\sigma}^{\text{unif}}}{\tau_{\sigma}},\ w_{\sigma}=\frac{t_{\sigma}-1}{t_{\sigma}+1}
\]
in all of the above functionals with the exceptions of M05, M06, M06-2X,
M06-SO, M06-L, revM06, and revM06-L. 

The present finite element study of these functionals is motivated
by the finding of \citet{Mardirossian2013_JCTC_4453} that many Minnesota
functionals converge slowly to the CBS limit; yet many of the presently
considered Minnesota functionals were published after \citeref{Mardirossian2013_JCTC_4453}.
We also found surprisingly large Gaussian basis set truncation errors
in exploratory FEM calculations with the M06-L\citep{Zhao2006_JCP_194101}
functional in a recent study.\citep{Schwalbe2022_JCP_174113}

\begin{table*}
\begin{centering}
\begin{tabular}{lccc}
Method & LIP & 1st order HIP & 2nd order HIP\tabularnewline
\hline 
\hline 
HF & -0.5000000 & -0.5000000 & -0.5000000\tabularnewline
TPSS & -0.5002355 & -0.5002355 & -0.5002355\tabularnewline
TASKCC & -0.5001730 & -0.5001730 & -0.5001730\tabularnewline
r$^{2}$SCAN & -0.5001732 & -0.5001732 & -0.5001732\tabularnewline
M05 & -0.4999407 & -0.4999432 & -0.4999372\tabularnewline
M05-2X & N/C & N/C & N/C\tabularnewline
M06 & -0.5012906 & -0.5012928 & -0.5012897\tabularnewline
M06-2X & N/C & N/C & N/C\tabularnewline
M06-SX & -0.4856891 & -0.4856891 & -0.4856891\tabularnewline
M06-L & -0.5049877 & -0.5049876 & -0.5049872\tabularnewline
revM06 & -0.4978698 & -0.4978698 & -0.4978698\tabularnewline
revM06-L & -0.5000720 & -0.5000720 & -0.5000720\tabularnewline
M08-SO & N/C & -0.5039712 & -0.5039475\tabularnewline
M08-HX & -0.5039980 & -0.5039981 & -0.5039979\tabularnewline
M11 & -0.4998235 & -0.4998251 & -0.4998216\tabularnewline
revM11 & -0.5023467 & -0.5023467 & -0.5023467\tabularnewline
M11-L & -0.5128008 & -0.5127346 & -0.5126698\tabularnewline
MN12-SX & -0.4970768 & -0.4970769 & -0.4970768\tabularnewline
MN12-L & -0.4923232 & -0.4923232 & -0.4923232\tabularnewline
MN15 & -0.4997453 & -0.4997453 & -0.4997453\tabularnewline
MN15-L & -0.4965988 & -0.4965988 & -0.4965988\tabularnewline
$\omega$B97X-noV & -0.5053272 & -0.5053272 & -0.5053272\tabularnewline
B97M-noV & -0.5061077 & -0.5061077 & -0.5061077\tabularnewline
$\omega$B97M-noV & -0.4992064 & -0.4992064 & -0.4992064\tabularnewline
\end{tabular}
\par\end{centering}
\caption{Total energies for H atom with various numerical basis sets. N/C:
Calculation did not reach SCF convergence in 500 iterations. \label{tab:Total-energies-for}}

\end{table*}

All total energies are summarized in \tabref{Total-energies-for}.
The TPSS, TASKCC or r$^{2}$SCAN functionals pose no issues for the
H atom, and converge to the same total energies in all three numerical
basis sets.

Continuing to the semiempirical Minnesota functionals, starting out
with M05, we observe that the results for M05 are not converged to
the CBS limit, as is also obvious from the different total energies
obtained in the various calculations in \tabref{Total-energies-for}.
The poor convergence is partly explained by strong oscillations observed
in the $\tau$ and $\nabla^{2}n$ of the solutions (see Supplementary
Information), which speak to the numerical ill-behavedness of the
functional. SCF calculations for M05-2X, in turn, fail in all basis
sets, suggesting numericall ill-behavedness also in this functional.

M06 is not converged to the CBS limit, and large oscillations in $\nabla^{2}n$
are again observed in the SCF solutions; the same observations also
apply to M06-L. M06-2X is too unstable to reach SCF convergence like
M05-2X. In contrast, the recent revM06, revM06-L, and M06-SX functionals
appear well-behaved: the calculations in all three basis sets converge
to the same total energy and $\tau$ and $\nabla^{2}n$ appear smooth.

M08-SO fails to yield a converged SCF solution in the LIP basis, and
as evidenced by the difference of the first-order and second-order
HIP results, is also not converged to the CBS limit; large oscillations
in $\nabla^{2}n$ are observed in the corresponding solutions. M08-HX
is almost at the CBS limit, as the total energies reproduced by the
three basis sets agree to within $0.2\mu E_{h}$. However, sharp non-physical
behavior is observed in $\nabla^{2}n$ of the solution.

M11 and M11-L are also characterized by lack of CBS convergence and
sharp oscillations in $\nabla^{2}n$. In contrast, revM11 is better
behaved, but features a bump and shoulder in $\nabla^{2}n$ that do
not exist in revM06 or revM06-L that are much closer to the exact
$\nabla^{2}n$.

MN12-SX, MN12-L, MN15, and MN15-L are well-behaved for the H atom:
all three basis sets yield the same total energy and the resulting
$\nabla^{2}n$ is smooth. MN12-L and MN12-SX feature a shoulder that
is similar to, but less pronounced than that for revM11.

All members of the Berkeley family are well-behaved: $\omega$B97X-noV,
B97M-noV and $\omega$B97M-noV all converge smoothly to the CBS limit
with the studied basis sets and show no sharp features in $\nabla^{2}n$,
although the two meta-GGAs reproduce $\nabla^{2}n$ which is slightly
different from the exact value.

\subsection{Potential implications for construction of NAO basis sets \label{subsec:nao}}

In this section, we investigate whether the control of derivatives
afforded by HIPs could be useful for NAO basis set construction. A
key difference between NAO basis sets and analytical basis sets commonly
used in LCAO discussed in the Introduction is that NAO basis sets
have finite support:\citep{Lehtola2019_IJQC_25968} the NAO basis
functions vanish outside a given cutoff radius $R$ from the nucleus---$\chi_{\alpha}(\boldsymbol{r})=0$
for $r>R$---which affords significant sparsity in large systems
that is commonly exploited for performance benefits. The established
approach to build NAO basis sets is to use FDM with various confinement
potentials,\citep{Sankey1989_PRB_3979,Porezag1995_PRB_12947,Horsfield1997_PRB_6594,SanchezPortal1997_IJQC_453,Kenny2000_PRB_4899,Junquera2001_PRB_235111,Anglada2002_PRB_205101,Ozaki2003_PRB_155108,Ozaki2004_JCP_10879,Ozaki2004_PRB_195113,Blum2009_CPC_2175,Shang2010_IRPC_665,Louwerse2012_PRB_35108,Corsetti2013_JPCM_435504}
as it is desirable that both the radial function (\eqref{fem-func})
and its derivative
\begin{equation}
\chi_{\mu}'(r)=-\frac{B_{\mu}(r)}{r^{2}}+\frac{B_{\mu}'(r)}{r}\label{eq:chimuder}
\end{equation}
go smoothly to zero when approaching the cutoff radius $R$, as large
derivatives $\chi{}_{\mu}'(r)$ close to the cutoff $r\approx R$
are problematic for the evaluation of molecular integrals by quadrature.

In FEM, the LIP basis allows direct control on the boundary value
$B_{\mu}(r_{\infty})$, which also appears in the first term of \eqref{chimuder}.
A key feature of HIPs is that they allow explicit control also on
the boundary conditions for derivative(s), such as the second term
of \eqref{chimuder}. We will therefore investigate whether the added
control on the second term provided by the HIP basis is useful for
building NAO basis sets by studying the magnesium atom with LIP and
HIP basis sets with various values for the practical infinity $r_{\infty}$.
In these calculations, we employ 5 radial elements with 15-node LIPs
or 8-node HIPs and the PBE functional.\citep{Perdew1996_PRL_3865,Perdew1997_PRL_1396}

\begin{figure}[H]
\begin{centering}
\includegraphics[width=1\linewidth]{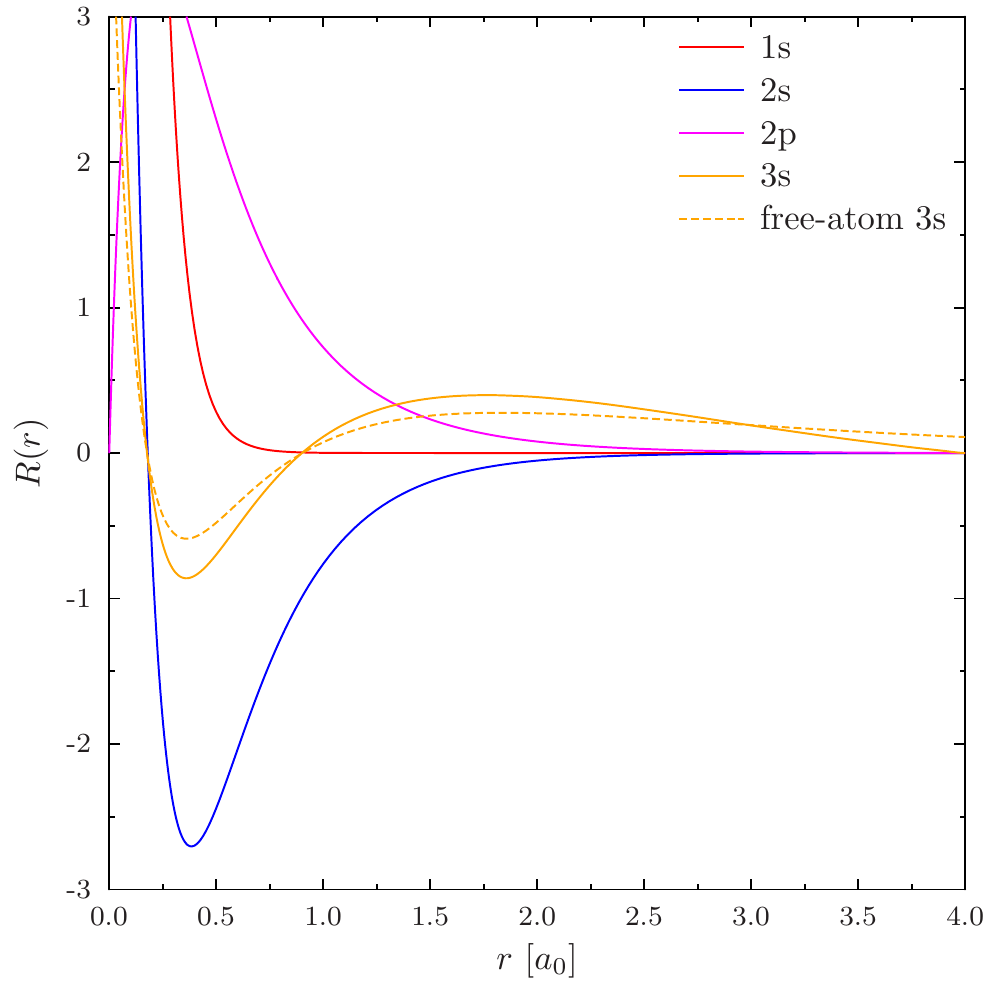}
\par\end{centering}
\caption{Radial orbitals of Mg with a 8-node HIP basis with 5 radial elements,
the PBE functional and $r_{\infty}=4a_{0}$. The free-atom $3s$ orbital
($r_{\infty}=40a_{0}$) is also shown as reference. The difference
between the solutions for $B_{\mu}'(r_{\infty})\protect\neq0$ and
$B_{\mu}'(r_{\infty})=0$ is too small to be seen in the plot. \label{fig:Mg-orbs}}
\end{figure}

\begin{figure}[H]
\begin{centering}
\includegraphics[width=1\linewidth]{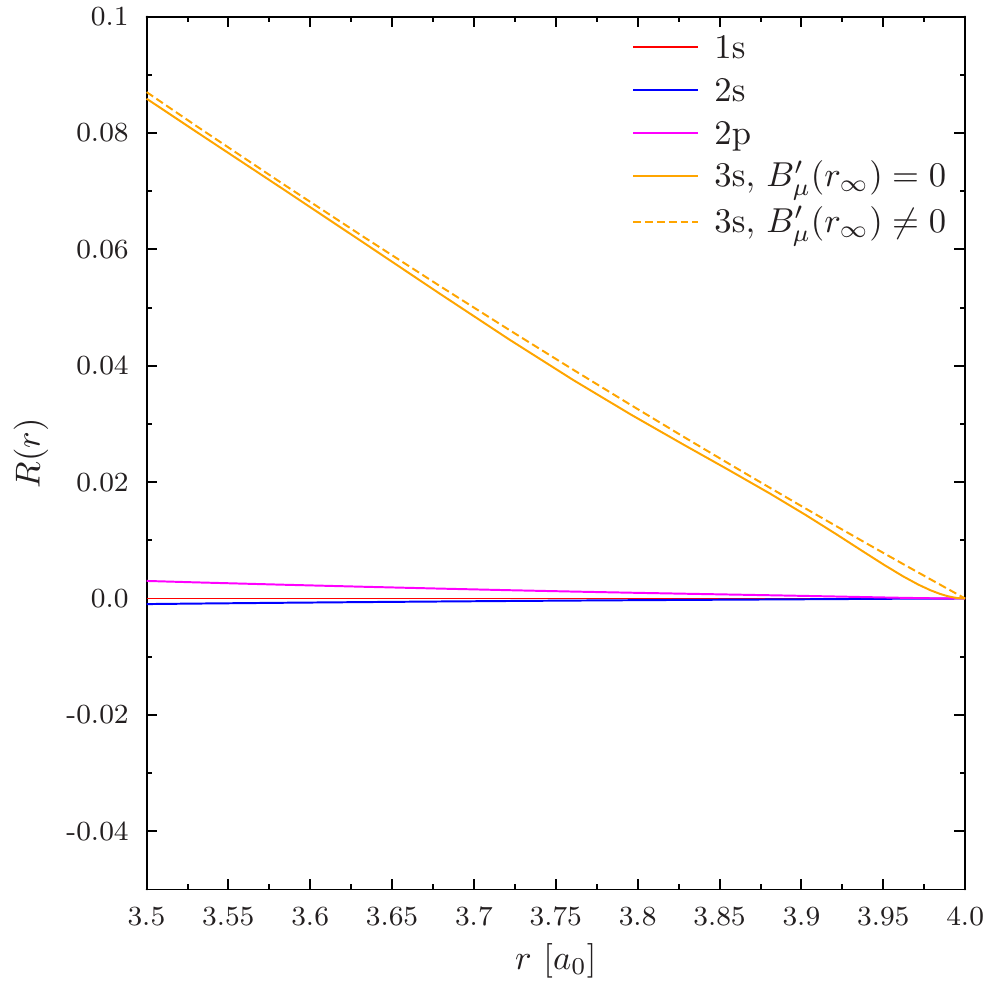}
\par\end{centering}
\caption{Closeup of radial orbitals of Mg shown in \figref{Mg-orbs}, focused
in the region near $r_{\infty}=4a_{0}$, difference between the solutions
for $B_{\mu}'(r_{\infty})\protect\neq0$ and $B_{\mu}'(r_{\infty})=0$
is visible. \label{fig:Mg-closeup}}
\end{figure}

In this demonstration, we will consider LIPs and two types of calculations
with HIPs: one where a finite derivative is allowed at $r_{\infty}$
(yielding analogous results to the LIP basis), and another one where
the derivative is forced to vanish at $r_{\infty}$. The value $r_{\infty}=4a_{0}$
suffices for this demonstration. The resulting radial 1s, 2s, 2p,
and 3s orbitals are shown in \figref{Mg-orbs}; only the 3s orbital
turns out to differ significantly from the free atom. The issue for
NAO basis set construction is that the 3s orbital develops a near-constant
slope for large $r$ in the constrained atom, all the way up to $r_{\infty}$.
The derivative can, however, be made to vanish at $r_{\infty}$ as
demonstrated by \figref{Mg-closeup}.

\begin{table}[H]
\begin{centering}
\begin{tabular}{cccc}
$N_{\text{elem}}$ & $N_{\text{bf}}$ & $E[B_{\mu}'(r_{\infty})\neq0]$ & $E[B_{\mu}'(r_{\infty})=0]$\tabularnewline
\hline 
\hline 
1 & 13 & -198.854894880 & -198.818748501\tabularnewline
2 & 27 & -199.616629182 & -199.610609917\tabularnewline
3 & 41 & -199.616629885 & -199.611532191\tabularnewline
4 & 55 & -199.616629940 & -199.612275491\tabularnewline
5 & 69 & -199.616629942 & -199.612844761\tabularnewline
10 & 139 & -199.616629942 & -199.614364754\tabularnewline
20 & 279 & -199.616629942 & -199.615382891\tabularnewline
40 & 559 & -199.616629942 & -199.615974661\tabularnewline
80 & 1119 & -199.616629942 & -199.616293874\tabularnewline
\end{tabular}
\par\end{centering}
\caption{Convergence of the PBE total energy of Mg with $r_{\infty}=4a_{0}$,
either allowing a finite derivative of the radial function at the
practical infinity, or disallowing it. The default finite element
grid with $z=2$ is employed. \label{tab:Convergence-of-the}}
\end{table}

But, as is obvious by the oscillations of the $B_{\mu}'(r_{\infty})=0$
curve in \figref{Mg-closeup}, this calculation is not converged to
the CBS limit even though the $B_{\mu}'(r_{\infty})\neq0$ calculation
is, as is demonstrated in \tabref{Convergence-of-the}. As these data
show, imposing the zero-derivative condition appears to make the calculation
sensitive to the finite element representation. 

\begin{figure}[H]
\begin{centering}
\includegraphics[width=1\linewidth]{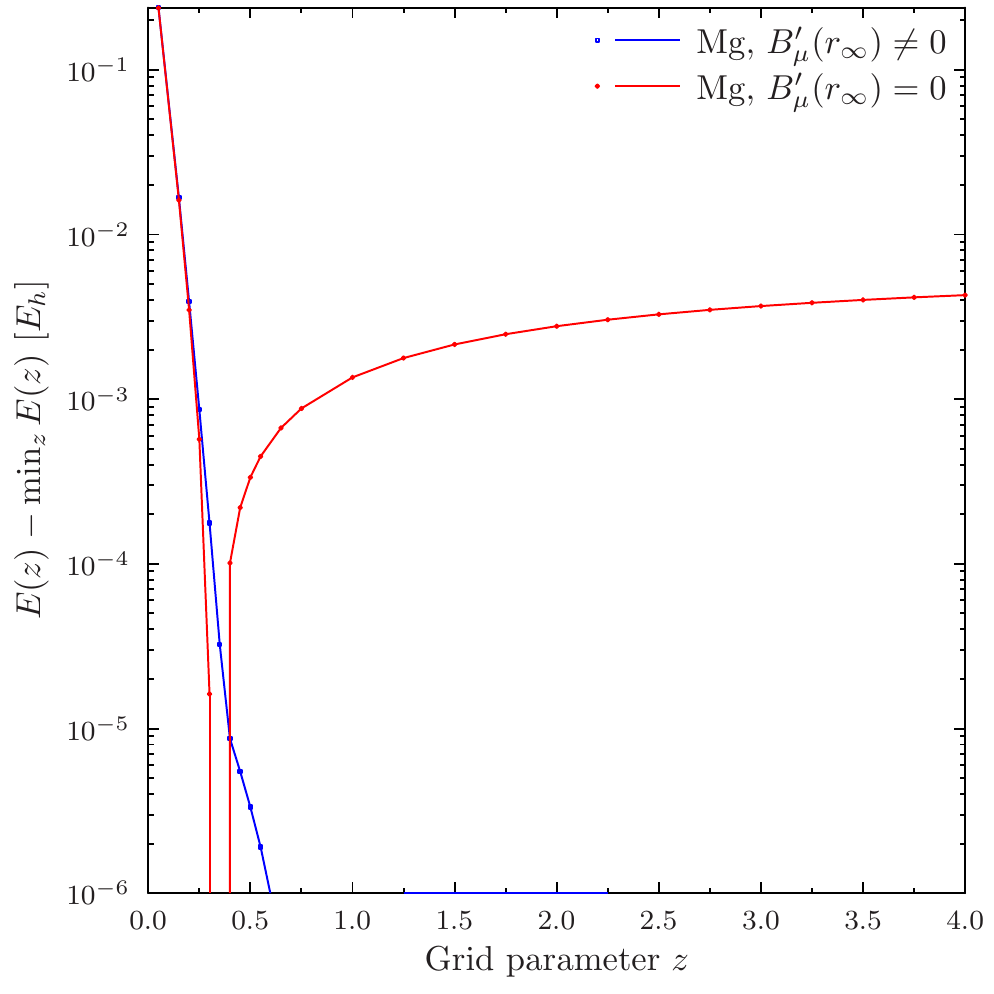}
\par\end{centering}
\caption{Dependence of the PBE total energy of Mg with $r_{\infty}=4a_{0}$
on the grid parameter $z$ in \eqref{expgrid} with five 8-node HIP
radial elements. Note logarithmic $y$ axis. \label{fig:Mg-z}}
\end{figure}

In any case, convergence of $E[B_{\mu}'(r_{\infty})=0]$ to the CBS
limit is slow, and it makes sense to ask whether the grid could be
improved. The element grid dependence is investigated in \figref{Mg-z}.
While the $B_{\mu}'(r_{\infty})\neq0$ calculation shows negligible
dependence on the value of the grid parameter $z$ (\eqref{expgrid}),
as the calculations with $z\geq0.6$ are converged to the sub-$\mu E_{h}$
accuracy, the $B_{\mu}'(r_{\infty})=0$ curve has a sharp minimum
around $z=0.33$, which is strikingly different from the recommended
value $z=2.0$ of \citeref{Lehtola2019_IJQC_25945}. This indicates
that the solution for $B_{\mu}'(r_{\infty})=0$ is sensitive to the
description of the wave function in the region near $r_{\infty}$,
as such a choice places much more elements in the far-valence region
than near the core, which furthermore suggests that CBS limit studies
are impractical for the $B_{\mu}'(r_{\infty})=0$ boundary condition.

Combined with the realization that $E[B_{\mu}'(r_{\infty})\neq0]$
is clearly an upper bound for $E[B_{\mu}'(r_{\infty})=0]$, since
the latter calculation has a constraint missing in the former one,
we are forced to conclude that the added control on the derivative
offered by the HIP basis does not appear to offer further benefits
over LIPs for building NAO basis sets.

Finally, because of the slow convergence seen in the $E[B_{\mu}'(r_{\infty})=0]$,
our advice is not to remove the function describing the derivative
at $r_{\infty}$ as was done in \subsecref{suphip} to make the LIP
and HIP calculations have exactly the same total number of radial
basis functions, since keeping the extra function is hugely important
for quick convergence in cases where $r_{\infty}$ is not converged
to the free-atom limit.

\section{Summary and Discussion \label{sec:Summary-and-Discussion}}

We have discussed how the numerical basis functions $\chi_{\mu}(r)=B_{\mu}(r)/r$
for atomic calculations can be evaluated in a numerically stable manner
at points close to the origin. When a Taylor expansion of a lower
polynomial order than that of $B_{\mu}(r)$ is employed, the error
made in the $d$th derivative at the switching point $\chi_{\mu}^{(d)}(R)$
increases with the order of the derivative $d$. However, if the order
of the Taylor expansion matches that of $B_{\mu}(r)$, the Taylor
series is accurate for all $d$ and we recommend the use of such small-$r$
expansions for all atomic calculations.

We have described the implementation of analytic first-order Hermite
interpolating polynomials (HIPs) and numerical general-order Hermite
interpolating polynomials, and studied their use in atomic electronic
structure calculations. We have shown that HIPs can successfully be
used in combination with a large number of nodes as well as with non-uniform
finite element grids, where they afford results that are as good as
or even better than those obtained with Lagrange interpolating polynomials
(LIPs) used in our previous works.\citep{Lehtola2019_IJQC_25944,Lehtola2019_IJQC_25945}
Furthermore, we demonstrated with the zinc atom that the grid dependence
of LDA, GGA and meta-GGA functionals is similar to that of Hartree--Fock
(HF) theory. (For a further application of 10-node HIPs, see \citeref{Lehtola2023__b}
on the study of a recently proposed analytic, regularized nuclear
Coulomb potential.)

We studied the importance of the HIP basis in meta-GGA calculations
on the hydrogen atom. Calculations in small numerical basis sets were
used to demonstrate that the LIP basis can yield a discontinuous local
kinetic energy density $\tau$ at element boundaries, while the first-order
HIP basis can reproduce kinks in $\tau$ already in calculations at
the HF level; only the second-order (or higher) HIP basis is guaranteed
to make $\tau$ smooth across element boundaries. We also identified
self-consistent field (SCF) convergence problems with otherwise well-behaved
$\tau$-dependent meta-GGA functionals if a small numerical basis
was employed, and explained this by the insufficient flexibility of
such numerical basis sets. Finally, we showed that given a sufficiently
large numerical basis set, all three choices for the shape functions
reproduced similar results for HF and $\tau$-dependent meta-GGA functionals
for the hydrogen atom. 

Our self-consistent calculations on hydrogen examined a large selection
of functionals, including TPSS, TASKCC, r$^{2}$SCAN, the whole Minnesota
family---M05, M05-2X, M06, M06-2X, M06-SX, M06-L, revM06, revM06-L,
M11, M11-L, revM11, MN12-SX, MN12-L, MN15, MN15-L---and the three
most recent Berkeley functionals $\omega$B97X-V, B97M-V, and $\omega$B97M-V
without non-local correlation. TPSS, TASKCC, r$^{2}$SCAN and all
three Berkeley functionals were found to converge without problems
in all three basis sets. M05, M05-2X, M06, M06-L, M06-2X, M08-SO,
M11, and M11-L failed to either reach SCF convergence, or the CBS
limit despite the use of a large finite element basis set. The observed
instabilities in these functionals are likely caused by oscillations
and/or large values in the functionals' enhancement factors.\citep{Mardirossian2013_JCTC_4453}
M08-HX was found to have converged close to the CBS limit, but it
was found to exhibit non-physical sharp features in $\nabla^{2}n$.
Only the most recent Minnesota functionals---the \citeyearpar{Wang2020_PNASUSA_2294}
M06-SX, the \citeyearpar{Verma2019_JPCA_2966} revM11, the \citeyearpar{Wang2018_PNASUSA_10257}
revM06, the \citeyearpar{Wang2017_PNASUSA_8487} revM06-L, and the
\citeyearpar{Yu2016_JCTC_1280} MN15 functional---as well as the
older \citeyearpar{Peverati2012_PCCP_13171} MN12-L and the \citeyearpar{Peverati2012_PCCP_16187}
MN12-SX functionals were found to converge without issues to the CBS
limit. Interestingly, even though the aforementioned newer functionals
share the same functional forms as the original ill-behaved parametrizations,
the revised parameter values in M06-SX, revM11, and revM06-L appear
to have removed the pathological behaviors in the earlier variants
of the M06 and M11 type of functionals. Even though this part of the
study was restricted to the hydrogen atom for simplicity, these results
are directly useful also for calculations on other systems given that
density functional approximations do not depend on the system. The
numerical properties of Minnesota functionals and other recent meta-GGAs
are studied on a wider variety of atoms in \citeref{Lehtola2023__a}. 

Although we have found that LIPs and HIPs work equally well for $\tau$-dependent
meta-GGAs when a large enough numerical basis set is used, calculations
with density Laplacian dependent functionals may have more stringent
requirements on the underlying shape function basis, as we exemplified
with small HF calculations on the hydrogen atom. Higher-than-first
order Hermite interpolating polynomials, which were studied numerically
in this work, would be relevant for such efforts. General analytical
formulas for such polynomials have been suggested in the literature,\citep{Spitzbart1960_AMM_42,ElZafrany1985_CANM_85,Wang2007_SiCSAM_1651}
but they do not appear to have been used in practice. Further work
should investigate the practical usefulness of the analytic expressions,
as it is not clear whether implementations thereof will be sufficiently
stable numerically.

\section*{Supporting Information}

Errors for approximate Taylor expansions from $3^{\text{rd}}$ to
$16^{\text{th}}$ order with a 15-node LIP basis and an 8-node analytical
first-order HIP basis. Plots of $\tau$ and $\nabla^{2}n$ for the
hydrogen atom for all studied functionals.

\section*{Acknowledgments}

I thank Dage Sundholm for discussions on numerical stability, and
the Academy of Finland for financial support under project numbers
350282 and 353749.
\begin{tocentry}
\includegraphics{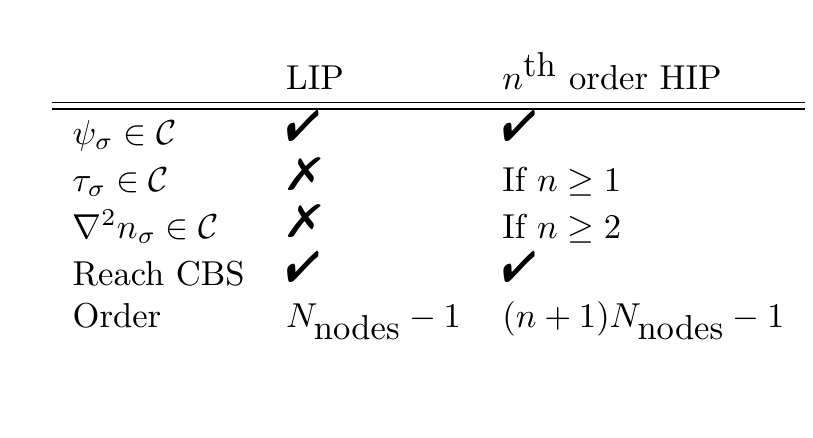}
\end{tocentry}
\bibliography{citations}

\begin{mcitethebibliography}{83}
\bibitem[Not()]{Note-1}
There are altogether $(n+1)N_{\text{nodes}}N_{\text{elem}}$ shape functions in
  the elements, out of which $(n+1)(N_{\text{elem}}-1)$ are at element
  boundaries and are therefore overlaid, and the first function value (not the
  derivatives!) is removed for the boundary condition at the nucleus, and the
  last function value and all $n$ derivatives are removed for the boundary
  conditions at the practical infinity.\relax
\mciteBstWouldAddEndPunctfalse
\mciteSetBstMidEndSepPunct{\mcitedefaultmidpunct}
{}{\mcitedefaultseppunct}\relax
\EndOfBibitem
\end{mcitethebibliography}

\end{document}